\documentclass[12pt,a4paper]{article}            
 \usepackage[skins,theorems]{tcolorbox}
\tcbset{highlight math style={enhanced,
  colframe=red,colback=white,arc=0pt,boxrule=1pt}}
  \usepackage[bookmarksopen, bookmarksnumbered, bookmarksopenlevel=2]{hyperref}
  \usepackage{tikz}
  \usepackage{tikz-3dplot}
 \usetikzlibrary{calc}
 \usetikzlibrary{decorations} %
 \usepackage[UKenglish]{babel}
 \usepackage[toc,page]{appendix}
 \usepackage{amsmath}
 \usepackage{amssymb}
 \usepackage{graphicx}
 \usepackage{hhline}
 \usepackage[bf]{caption}
\usepackage{cite}
\usepackage[vcentermath]{youngtab}
\usepackage{geometry}
\usepackage{slashed}
\usepackage{color}
\usepackage{stackrel}
\usepackage{tikz-cd} 
\usepackage{mathtools}
\usepackage{cancel} 
\usepackage{multirow}

\usepackage{subcaption}
\usepackage{empheq}
\usepackage{arydshln}

\newcommand{\bqa}{\begin{eqnarray}}
\newcommand{\eqa}{\end{eqnarray}}

\newenvironment{eqn*}{\begin{equation*}\begin{aligned}}{\end{aligned}\end{equation*}\noindent}
\hypersetup{
    pdftitle={},
    pdfauthor={},
    pdfsubject={}
}
\numberwithin{equation}{section}
\numberwithin{table}{section}

\makeatletter

\newcommand{\be}{\begin{equation}}
\newcommand{\ee}{\end{equation}}
\newcommand{\beq}{\begin{equation}}
\newcommand{\eeq}{\end{equation}}
\newcommand{\ba}{\begin{aligned}}
\newcommand{\ea}{\end{aligned}}

\newcommand{\bea}{\begin{eqnarray}}
\newcommand{\eea}{\end{eqnarray}}

\newcommand{\cT}{\mathcal{T}}

\newcommand{\cI}{\mathcal{I}}
\newcommand{\cJ}{\mathcal{J}}

\newcommand{\cV}{\mathcal{V}}

\newcommand{\cM}{\mathcal M}

\newcommand\bi{\begin{itemize}}
\newcommand\ei{\end{itemize}}

\def\unit{{1\kern-.65ex {\rm l}}}
\def\1{{1\kern-.65ex {\rm l}}}

\def\bbP{\mathbb{P}}

\def\bbH{{\mathbb{H}}}
\def\bbI{{\mathbb{I}}}

\def\bbN{{\mathbb{N}}}

\def\bbP{{\mathbb{P}}}
\def\bbQ{{\mathbb{Q}}}
\def\bbR{{\mathbb{R}}}

\def\bbZ{{\mathbb{Z}}}

\newcount\hour \newcount\minute
\hour=\time \divide \hour by 60
\minute=\time
\def\now{%
\ifnum \hour<13
  \ifnum \hour=0 \advance \hour by 12 \number\hour:\else \number\hour:\fi%
     \ifnum \minute<10 0\fi%
     \number\minute%
\ A.M.%
\else \advance \hour by -12 \number\hour:%
  \ifnum \minute<10 0\fi%
  \number\minute%
  \ P.M.%
\fi%
}

\makeatother

\begin{document}

\begin{titlepage}
\begin{center}
\rightline{\small }

\vskip 15 mm

{\large \bf

} 
\vskip 11 mm

\begin{center}
{\Large \bfseries Charting the Complex Structure Landscape of F-theory}\\[.25em]

\vspace{1cm}
{\bf Damian van de Heisteeg}

\vskip 11 mm

{\small \it
Center of Mathematical Sciences and Applications \& Jefferson Physical Laboratory, \\
Harvard University, Cambridge, MA 02138, USA\\[3mm]
}

\vspace*{1.5em}

\end{center}

\vskip 11 mm

\end{center}
\vskip 17mm

\begin{abstract}
\noindent 
We explore the landscape of F-theory compactifications on Calabi--Yau fourfolds whose complex structure moduli space is the thrice-punctured sphere. As a first part, we enumerate all such Calabi--Yau fourfolds under the additional requirement that it has a large complex structure and conifold point at two of the punctures. We find 14 monodromy tuples by demanding the monodromy around infinity to be quasi-unipotent. As second part, we study the four different types of phases arising at infinity. For each we consider a working example where we determine the leading periods and other physical couplings. We also included a notebook that sets up the period vectors for any of these models.
\end{abstract}

\vfill

\vspace*{13em}

{\ \ \  \small \verb "dvandeheisteeg@fas.harvard.edu" }

\end{titlepage}

\newpage

\tableofcontents
\newpage

\section{Introduction}
Compactifications of string theory result in a vast landscape of lower-dimensional field theories, featuring scalar field spaces with varying physical couplings. To assess the validity of these effective descriptions, one needs to know what phases can occur within these field spaces, and identify the effects that could trigger a breakdown. Consequently, we can think of the landscape of all field spaces as emerging from how different phases can be put together. Both improving our understanding of these limits in field space, as well as mapping out the landscape of possible effective field theories, are central to the Swampland program \cite{Vafa:2005ui} (see \cite{Palti:2019pca, vanBeest:2021lhn,Agmon:2022thq} for reviews).

Limits in field spaces are of natural interest to the Distance Conjecture \cite{Ooguri:2006in}, which posits that as we traverse towards asymptotic regions at infinite distance, infinite towers of states should become light. Recently this idea has been scrutinized meticulously, particularly in Type II and M-theory Calabi--Yau compactifications \cite{Grimm:2018ohb, Blumenhagen:2018nts, Lee:2018urn, Grimm:2018cpv, Corvilain:2018lgw, Joshi:2019nzi, Font:2019cxq, Marchesano:2019ifh,  Grimm:2019wtx,Erkinger:2019umg, Lee:2019oct, Baume:2019sry, Klawer:2021ltm, Alvarez-Garcia:2021mzv, Alvarez-Garcia:2021pxo}, by studying the scalar field spaces spanned by the K\"ahler and complex structure moduli. However, the phases arising in these moduli spaces are not just limited to infinite distance regimes, as they also encompass finite distance loci such as conifold or Landau-Ginzburg points, cf.~\cite{Candelas:1990rm}. In order to get a complete picture, it is therefore important to consider these finite distance phases as well.

The vastness of the string landscape comes partially from the large number of Calabi--Yau manifolds that we can use as compactification geometry. In fact, even finiteness of Calabi--Yau manifolds is still a major open question, and has only been answered affirmatively for certain classes \cite{gross1993finiteness, wilson2023boundedness}. When focusing on certain databases such as Kreuzer-Skarke \cite{Kreuzer:2000xy}, explicit countings have been performed recently in \cite{Gendler:2023ujl, Chandra:2023afu}. In contrast, in this paper we take a different approach that does not rely on the geometry itself: we fix a particular moduli space and determine what sets of coupling functions are compatible with it. 

The focus of this paper is a very particular corner of the landscape: Calabi--Yau fourfolds whose complex structure moduli space is the thrice-punctured sphere, $\cM_{\rm cs} = \bbP^1\backslash\{0,1,\infty\}$. From this class we want to draw lessons about how this landscape comes about and what phases can arise within it. We address the first by enumerating all such Calabi--Yau fourfolds under some requirements on the singularity types, finding 14 manifolds in the process. For the second we study the phases occurring at the singularity at infinity, for which we write down physical couplings such as K\"ahler potentials and flux superpotentials.

Calabi--Yau manifolds with $\cM_{\rm cs} = \bbP^1\backslash\{0,1,\infty\}$ have served as a helpful guide in understanding moduli spaces arising in string compactifications. In fact, the mirror quintic threefold studied in the seminal work \cite{Candelas:1990rm} on mirror symmetry is a prime example. One naturally wonders whether there are more Calabi--Yau threefolds of this type, and this question was ultimately answered in \cite{Doran:2005gu}, resulting in fourteen cases in total. On the other hand, for Calabi--Yau fourfolds a similar classification has been missing so far, although nine of them were readily examined by \cite{CaboBizet:2014ovf} in the study of F-theory compactifications.

Provided the Calabi--Yau fourfolds are elliptically fibered, their complex structure moduli spaces encode distinctive aspects about F-theory compactifications: in addition to the usual complex structure moduli of the closed string sector, the complex structure deformations of the fourfold also describe the axio-dilaton and open string moduli. Recently, this enabled non-trivial tests of the distance conjecture in F-theory compactifications to eight and six dimensions on Calabi--Yau threefolds and K3 surfaces \cite{Lee:2021qkx,Lee:2021usk,Alvarez-Garcia:2023gdd,Alvarez-Garcia:2023qqj}. Our Calabi--Yau fourfolds are elliptic fibrations over $\bbP^1\times \bbP^1 \times \bbP^1$, thereby giving us a glimpse into the landscape of four-dimensional $\mathcal{N}=1$ F-theory compactifications. And even though we have only a single complex structure modulus, the non-geometric phases at infinity still exhibit some effects characteristic of F-theory.

An appropriate framework for constraining the global structure of the moduli space is given by the monodromy group $\Gamma$. As we move through the moduli space and wind around singularities, our physical couplings undergo duality transformations, altogether generating $\Gamma$. Since these couplings are encoded by period integrals associated to the Calabi--Yau manifolds, it follows that properties of these functions such as holomorphicity may be formalized into conditions on $\Gamma$ \cite{Griffiths1970}. These conditions explain for instance that moduli spaces with only one or two punctures must have constant coupling functions. There is even a finiteness theorem \cite{DeligneFiniteness}, stating that there are only finitely many monodromy groups for a given moduli space with fixed singularity structure, but unfortunately the proof does not give an effective method for enumerating them. In turn, given a monodromy group $\Gamma$, the period vector that gives rise to this monodromy behavior is determined uniquely, which follows from the rigidity theorem of \cite{Schmid} (see \cite{Doran:2005gu} for the precise argument for Calabi--Yau manifolds with $\cM_{\rm cs} = \bbP^1\backslash\{0,1,\infty\}$). 

An effective approach for enumerating monodromy groups on $\cM_{\rm cs}=\bbP^1\backslash\{0,1,\infty\}$ was laid out in \cite{Doran:2005gu} by Doran and Morgan, where all Calabi--Yau threefolds with this moduli space were classified. This classification was subject to an additional assumption about two of the singular points, requiring large complex structure and conifold points with the form of the monodromies constrained by mirror symmetry. We make similar assumptions in the fourfold case, giving us two monodromy matrices parametrized by the intersection number $\kappa$ and integrated second Chern class $c_2$ of the mirror. A priori $\kappa$ and $c_2$ may take infinitely many values, but the key insight of \cite{Doran:2005gu} was that monodromies around a singularity must be \textit{quasi-unipotent} \cite{Landman, Schmid}: some finite power of the monodromy matrix $M$ is unipotent, i.e.~$(M^l-1)^{d+1}=0$ for some $d,l$. Demanding quasi-unipotence of the monodromy $M_\infty$ around infinity, we find 14 possible values for $\kappa, c_2$ as summarized in table \ref{table:monodromydata}. The associated Calabi--Yau fourfolds are in one-to-one correspondence to the 14 Calabi--Yau threefolds of \cite{Doran:2005gu}. 

The phases arising at infinity can then be sorted into four types based on $M_\infty$: another large complex structure point, a CY3-point, a conifold point, and a Landau--Ginzburg point. Most of these types are already familiar from one-parameter Calabi--Yau threefolds \cite{almkvist2005tables, vanstraten2017calabiyau}, with CY3-points replacing K-points. From an F-theory point of view these CY3-points provide us with weak-coupling limits to Type IIB orientifolds on rigid Calabi--Yau threefolds. Furthermore, explicit control over the fourfold periods allows us to compute effects such as D($-1$)-instantons and D7-brane flux superpotentials. For each of these phases we present a working example giving for instance the leading periods, providing similar models as obtained in \cite{Bastian:2021hpc, Bastian:2023shf} for Calabi--Yau threefolds.

This paper is structured as follows. In section \ref{sec:review} we review some generalities about periods of Calabi--Yau manifolds and how to compute these functions in explicit examples. In section \ref{sec:classification} we classify the Calabi--Yau fourfolds with $\cM = \bbP^1\backslash\{0,1,\infty\}$ through their monodromy matrices. We enumerate all monodromy tuples, write down and solve the Picard-Fuchs equation, and give the corresponding mirror Calabi--Yau fourfolds. In section \ref{sec:infmonodromies} we then take a working example for each of the four possible phases at infinity, writing down leading periods and related physical couplings. In section \ref{sec:conclusions} we summarize our findings and point out some future directions. In appendix \ref{app:Ktheory} we relate our integral period basis to another that is commonly used in the literature. In appendix \ref{app:threefolds} we summarize all Calabi--Yau threefolds with $\cM_{\rm cs} = \bbP^1\backslash\{0,1,\infty\}$, and finally in appendix \ref{app:lmhs} we review some definitions on limiting mixed Hodge structures. 

As ancillary files we have included notebooks concerning the classification of monodromy tuples and the computation of Calabi--Yau periods. The notebooks about periods have been set up in a pedagogical manner for the reader interested in learning these techniques.

\section{Review on Calabi--Yau Periods}\label{sec:review}
Here we review the necessary background on periods of Calabi--Yau manifolds. In subsection \ref{ssec:CYperiods} we lay out the general structure of period vectors. The next subsection \ref{ssec:localperiods} gives a pedagogical introduction on how to compute these periods locally, and how to fix the integral basis and transfer it to other patches. In the final subsection \ref{ssec:T2} we illustrate our methods on the simple case of $T^2$. For the reader interested in learning these techniques we also recommend looking at the attached notebooks.

\subsection{Calabi--Yau periods and monodromies}\label{ssec:CYperiods}
We begin by giving a short review of the periods of Calabi--Yau manifolds. We put a particular emphasis on their monodromy matrices and the properties they satisfy; their so-called \textit{quasi-unipotence} will be one of our main tools in enumerating the landscape of moduli spaces later.

\paragraph{Periods and pairing.} In this work we are interested in the complex structure moduli space $\cM_{\rm cs}$ of Calabi--Yau manifolds $Y_D$, where $D$ denotes its complex dimension. In general this moduli space is $h^{D-1,1}$-dimensional, although in this paper we restrict to $h^{D-1,1}=1$. Many of the physical couplings arising in the string theory compactifications on these manifolds may be computed from so-called \textit{period integrals}, or periods for short. Denoting the (up to rescaling) unique holomorphic $(D,0)$-form  by $\Omega \in H^{D,0}$, we can expand it in terms of these periods as
\begin{equation}
    \Omega(z) = \Pi^{\cI}(z) \gamma_\cI \, ,\qquad \cI = 1, \ldots, \dim H^D\, ,
\end{equation}
where $\gamma_\cI \in H_D(Y_D,\bbZ)$ denotes an integral basis of $D$-forms. The periods $\Pi^{\cI}(z)$ combine into the period vector $\mathbf{\Pi} = (\Pi^{\cI})_{\cI=1,\ldots,\dim H^D}$. The intersection pairing for the basis $\gamma_\cI$ is denoted by
\begin{equation}
\eta_{\cI \cJ} = \int_{Y_D} \gamma_{\cI}\wedge\gamma_{\cJ}\, .
\end{equation}
This pairing is (skew)-symmetric depending on the parity of $D$: $\eta_{\cJ\cI} = (-1)^{D} \eta_{\cI\cJ}$. For Calabi--Yau threefolds its isometry group is $G=\text{Sp}(2h^{2,1}+2)$, while for Calabi--Yau fourfolds it is $G=\text{SO}(2+h^{2,2},2h^{3,1})$. 

\paragraph{Physical couplings.} The dependence of physical couplings arising from Calabi--Yau compactifications is determined by these period integrals. For example, for Type IIB on Calabi--Yau threefolds the $\mathcal{N}=2$ vector multiplet sector is determined by periods depending on the complex structure moduli. This paper concerns periods of elliptically fibered Calabi--Yau fourfolds, which capture (parts of) the $\mathcal{N}=1$ K\"ahler potential and flux superpotential in F-theory compactifications (see \cite{Haack:2001jz,Denef:2008wq, Grimm:2010ks,  Weigand:2018rez} for a discussion on M/F-theory duality). The K\"ahler potential for the complex structure moduli of a Calabi--Yau $D$-fold reads
\begin{equation}
    K_{\rm cs}(z,\bar z) = -\log i^D \int_{Y_D}\bar\Omega \wedge \Omega = -\log i^D \mathbf{\bar \Pi}^T \eta \mathbf{\Pi}\, .
\end{equation}
Similarly, for F-theory compactifications on Calabi--Yau fourfolds we may write the Gukov-Vafa-Witten flux superpotential \cite{Gukov:1999ya} in terms of the periods. The superpotential induced by turning on four-form flux $G_4 \in H^4(Y_4, \bbZ)$ reads \cite{Gukov:1999ya, Haack:2001jz,Grimm:2010ks}
\begin{equation}
    W = \int_{Y_4} G_4 \wedge \Omega = \mathbf{G}_4^T \eta \mathbf{\Pi}\, ,
\end{equation}
where $\mathbf{G}_4=(G_4^{\ \cI})_{\cI = 1,\ldots,\dim H^4_{\rm prim}}$. We ignore any non-perturbative terms for the K\"ahler moduli in this work. The scalar potential is then given by
\begin{equation}
    V = e^{K} K^{I\bar J}D_I W D_{\bar J}\overline{W}\, ,
\end{equation}
where $I,J = 1,\ldots, h^{3,1}$ run over the complex structure moduli $z^I$, $K_{I\bar{J}}$ denotes the K\"ahler metric and $D_I W=\partial_I W + (\partial_I K) W$ the K\"ahler covariant derivative. The dependence on the complex structure moduli is completely determined by the fourfold periods. We suppressed the dependence on the K\"ahler moduli, but it enters through the K\"ahler potential $K = K_{\rm cs}-2\log \cV_b$ via the base volume $\cV_b$.

\paragraph{Monodromies.} Next we consider the behavior of the periods around the singular loci in the moduli space. Let us consider such a singular point located at $z=0$. As we traverse around $z=0$, the period vector picks up a monodromy transformation $M \in G_\bbZ$, i.e.~some rotation by an integral matrix in Sp$(2h^{2,1}+2,\bbZ)$ or SO$(2+h^{2,2},2h^{3,1};\bbZ)$ as
\begin{equation}
    \Pi(e^{2\pi i }z ) = M\,  \Pi(z)\, .
\end{equation}
This monodromy is not any integral matrix in $G$, but it must also be \textit{quasi-unipotent} \cite{Schmid, Landman} (see Lemma 4.5 of \cite{Schmid} for an instructive group-theoretic proof). This means that for some finite order $l$ and nilpotency degree $d$ it satisfies
\begin{equation}\label{eq:quproperty}
        (M^l-\mathbb{I})^{d} \neq 0\, , \qquad         (M^l-\mathbb{I})^{d+1} = 0\, ,
\end{equation}
where $\bbI$ denotes the identity matrix. This quasi-unipotence property \eqref{eq:quproperty} plays a crucial part when we enumerate monodromy groups later in this work. We can perform a Jordan decomposition of the monodromy $M$ into two commuting factors $M_u,M_s \in G_\bbQ$ as
\begin{equation}
    M = M_u M_s\, ,
\end{equation}
where $M_u$ is its unipotent and $M_s$ its semisimple part. From \eqref{eq:quproperty} we then learn that $M_s$ must be of finite order, while $M_u$ must be of unipotency degree $d$, i.e.
\begin{equation}\label{eq:propertiesMsMu}
(M_s)^l = \mathbb{I}\, , \qquad \quad M_u = e^N : \quad N^d \neq 0\, , \ N^{d+1}=0\, .
\end{equation}
By going to a finite cover $z \to z^l$ we can remove the semisimple factor if we want to do so, as this would make us work instead with $M^l = (M_u)^l$. 

\paragraph{Minimal number of singular points.} In this work we focus on the thrice-punctured sphere  $\bbP^1\backslash\{0,1,\infty\}$ as complex structure moduli space. It is instructive to briefly review the argument that a lower number of punctures automatically results in a period vector that is constant. The two main results we use follow from theorem (9.8) of \cite{Griffiths1970}: (1) a finite monodromy group $\Gamma$ implies that the periods are constant,  and (2) the monodromy group $\Gamma$ must be completely reducible. Property (1) directly rules out the once-punctured sphere: a loop around the puncture is contractible around the other side of the sphere, so the monodromy must be trivial. For the twice-punctured sphere we need to use property (2). Recall that a completely reducible group is defined as a group that is a direct sum of irreducible representations. An example of a monodromy group that is \textit{not} completely reducible is given by the unipotent group\footnote{This group leaves the vector $(1,0)$ invariant, so the one-dimensional vector space spanned by $(1,0)$ gives an irreducible representation. However, there is no second irreducible representation, as $\Gamma_{\rm uni}$ does not have a second eigenvector, so we find that $\Gamma_{\rm uni}$ is not completely reducible.}
\begin{equation}\label{eq:Guni}
    \Gamma_{\rm uni}  = \left\{ \begin{pmatrix}
        1 & n \\
        0 & 1 
    \end{pmatrix} \ | \  n \in \bbZ \right\} \subseteq SL(2,\bbZ)\, .
\end{equation}
Note that this rules out a twice-punctured sphere with monodromies given by $n=\pm 1$. This no-go result can be extended straightforwardly to any twice-punctured sphere with infinite order monodromies, as we can always bring it into such a Jordan normal form. So we are left with finite order monodromy groups: these can be completely reducible, take for instance
\begin{equation}
    \Gamma_{\rm finite} = \left\{ \begin{pmatrix}
        1 & 0 \\
        0 & 1 
    \end{pmatrix} , \begin{pmatrix}
        0 & 1 \\
        -1 & 0 
    \end{pmatrix}  \right\}\, ,
\end{equation}
which has $(1,1)$ and $(1,-1)$ as basis vectors for its irreducible representations. However, from property (1) we know that periods corresponding to finite monodromy groups must be constant. So in conclusion, this leaves the thrice-punctured sphere $\bbP^1\backslash\{0,1,\infty\}$ as simplest non-trivial moduli space.

\subsection{Local periods and analytic continuation}\label{ssec:localperiods}
In this section we lay out the required techniques for computing periods in the various patches of moduli space. We explain how to set up the local solution for the periods near any singularity, and also describe how to compute the numerical transition matrices between two different local expansions. We aim to give a brief but pedagogical introduction to this subject, and recommend the reader interested in learning these techniques to explore the accompanying notebook.

\paragraph{Picard-Fuchs equation.} The periods of the holomorphic $(D,0)$-form of a Calabi--Yau manifold obey a set of differential equations known as the Picard-Fuchs system. We can solve these equations order-by-order for the periods by expanding around singular points in the moduli space. In practice, these differential systems are obtained from generalized Gel'fand-Kapranov-Zelevinsky (GKZ) hypergeometric systems \cite{GKZ} specified by the toric or complete intersection data of the Calabi--Yau manifold \cite{Hosono:1993qy, Hosono:1994ax}. In this work we specialize to the one-parameter case $h^{D-1,1}=1$, resulting in only a single Picard-Fuchs equation. Moreover, it suffices to consider those of ordinary hypergeometric type (see \cite{Norland} for a review on such differential equations)
\begin{equation}\label{eq:pfgen}
    L = \theta^{D+1} - \mu z \prod_{i=1}^{D+1} (\theta+a_i)\, , \qquad \theta = z \frac{d}{dz}
\end{equation}
where $z$ is the complex structure coordinate. The coefficients $a_i$ are rational numbers that depend on the model under consideration. We can read off the singularity discriminant as the coefficient of the highest monomial $\theta^{D+1}$, giving us in addition to $z=0,\infty$ another singularity at $1-\mu z =0$. By rescalings we are free to move it to any value; a convenient one is given by
\begin{equation}
    \mu^{-1} = e^{(D+1)\gamma_E+\sum_{i=1}^{D+1} \psi(a_i)}\, ,
\end{equation}
where $\gamma_E$ denotes the Euler-Mascheroni constant and $\psi(x)$ the digamma function. Remarkably, despite this intricate expression, in all examples $\mu$ will be an integer, see tables \ref{table:T2}, \ref{table:data} and \ref{table:hypergeom}. The sum of the coefficients $a_i$ satisfies
\begin{equation}
    \sum_{i=1}^{D+1} a_i = \frac{1}{2}(2D+1)\,.
\end{equation}
Moreover, we take $0 < a_i < 1$ and assume they are listed in increasing order.

\paragraph{Riemann symbol.} We can give a local solution for the periods by using the Frobenius method and expanding around a singularity. The exponents of the solutions of \eqref{eq:pfgen} at the singular points are conveniently summarized by the Riemann symbol
\begin{equation}\label{eq:P}
{\cal  P} \left\{\begin{array}{ccc}
0& 1/\mu& \infty\\ \hline
0& 0 & a_1\\
0& 1 & a_2\\
\vdots & \vdots & \vdots\\
0& D-1 & a_D\\
0& (D-1)/2 & a_{D+1}
\end{array}\right\}\, .
\end{equation}
The point $z=0$ is known as the large complex structure point, $z=1/\mu$ as the conifold point, and the type of singularity at $z=\infty$ depends on the coefficients $a_1,\ldots,a_{D+1}$. In contrast, at a regular point in moduli space the local exponents are $0,1,\ldots, D$. While the following discussion focuses on setting up local periods focuses on singular points, we note that the same strategy applies to regular points as well.

\paragraph{Frobenius solution.} Here we write down the general Frobenius solution for the periods. Let us fix a particular exponent $a$ and denote its multiplicity $m$. We then iterate over all pairs $(a,m)$ and construct a set of Frobenius solutions for each of them. The leading part of the $m$ periods is given by
\begin{equation}
    \varpi_{a,i}(z) = z^a\frac{\log[z]^i}{(2\pi i)^i}+\mathcal{O}(z^{a+1}) \, ,\qquad i=0,1,\ldots,m-1\, .
\end{equation}
We can parametrize the corrections to these periods explicitly in terms of power series. To illustrate this, let us first for simplicity consider the case where $m=2$. This corresponds to two periods, given to leading order by $\varpi_{a,0}(z)=z^a$ and $\varpi_{a,1}(z)=z^a\log[z]/2\pi i$. Then our ansatz for the expansions of the periods takes the form
\begin{equation}\label{eq:frobm=2}
    \varpi_{a,0}(z)=z^a f_{a,0}(z)\, , \qquad 2\pi i \,  \varpi_{a,1} = z^a f_{a,0}(z) \log[z]+z^a f_{a,1}(z)\, ,
\end{equation}
For the first period we consider a holomorphic function $f_{a,0}$, where we demand $f_{a,0}(0)=1$ to reproduce the leading term. For the second period we re-use the first function as coefficient of the logarithm, and add in a new holomorphic function $f_{a,1}$ satisfying $f_{a,1}(0)=0$. These holomorphic functions may be expanded explicitly as power series of the form given in \eqref{eq:powerseries}. We can determine their series coefficients order-by-order by plugging these ans\"atze into the Picard-Fuchs equation. Having gained some intuition from case $m=2$, we can repeat this pattern and build a general ansatz for all periods $\varpi_{a,0},\ldots,\varpi_{a,m-1}$ as 
\begin{equation}\label{eq:frobgen}
    (2\pi i)^i\varpi_{a,i}(z) = z^a \sum_{j=0}^{i}\binom{i}{j}
    f_{a,i-j}\log[z]^{j}\, , \qquad i=0,\ldots,m-1\, ,
\end{equation}
where the functions $f_{a,i}(z)$ are power series given by
\begin{equation}\label{eq:powerseries}
    f_{a,i} = \delta_{i,0}+\sum_{k=1}^\infty c_{a,i,k} z^k\, .
\end{equation}
Let us briefly elaborate on the structure in this Frobenius ansatz \eqref{eq:frobgen}:
\begin{itemize}
    \item The period $\varpi_{a,i}$ is made up of terms $f_{a,0}\log[z]^i, f_{a,1}\log[z]^{i-1}$, $\ldots$, through $f_{a,i}$. This generalizes how $\varpi_{a,1}$ above was made up of $f_{a,0}\log[z]$ and $f_{a,1}$. The overall factors of $(2\pi i )^i$ are included for integral monodromies under $z \to e^{2\pi i} z$.
    \item An added feature is the inclusion of binomial coefficients, which is typically referred to as the \textit{arithmetic} basis; for the purposes of this paper it is not very important, but for other applications it further refines the normalization of the rational coefficients $c_{a,i,k}$.
    \item  The $c_{a,i,k}$ may be obtained by plugging the ansatz \eqref{eq:frobgen} into the Picard-Fuchs equation \eqref{eq:pfgen} and solving order-by-order. Let us remark here that some of these coefficients may be left unfixed due to the freedom to take linear combinations of solutions; for instance, $\varpi_{a,0}=x^a+c_{a,0,1} x^{a+1}$ may be shifted by $\varpi_{a+1,0}=x^{a+1}$ to set $c_{a,0,1}=0$. We fix these ambiguities on an ad hoc basis, and clearly indicate how we do so.
\end{itemize} 
The above construction of the Frobenius solution applies to any one-modulus singularity, and we will specialize these ans\"atze to the particular sets of exponents we encounter for $T^2$ in section \ref{ssec:T2}, Calabi--Yau fourfolds in \ref{ssec:CY4localperiods} and Calabi--Yau threefolds in \ref{app:threefolds}.

\paragraph{Transition matrices.} The above formalism allows us to build local solutions around any point in moduli space up to some desired expansion order. These expansions are valid only within a certain patch of the moduli space, with the radius of convergence typically set by the nearest singularity. To move between two overlapping patches $A$ and $B$, we therefore need to compute transition matrices between these series solutions. Let us denote the two local Frobenius solutions by $\Pi_A$ and $\Pi_B$. Then we can expand both around some midpoint $x_{\rm mid} \in A\cap B$ in their overlap region as
\begin{equation}
    \Pi_{A,B}(x) = \sum_{k=0}^{\infty} \frac{1}{k!} \partial^k \Pi_{A,B}\big|_{x_{\rm mid}} (x-x_{\rm mid})^k\, .
\end{equation}
Taking the first $D+1$ terms, this gives us two complete bases $\partial^k \Pi_{A,B}$ (with $k=0,\ldots,D$) for the vector space. If we were able to evaluate both sets of period derivatives exactly at $x=x_{\rm mid}$, we would be able to give a closed form for the transition matrix. In practice, however, we take the series solutions computed up to sufficiently high order and evaluate these period derivatives numerically at $x=x_{\rm mid}$ for both $A$ and $B$. This yields a numerical transition matrix
\begin{equation}
    \mathcal{T}_A^{\; B} = \begin{pmatrix}
        \partial^0 \Pi_A & \cdots & \partial^{D} \Pi_A
    \end{pmatrix} \big|_{x_{\rm mid}} \begin{pmatrix}
        \partial^0 \Pi_B & \cdots & \partial^{D} \Pi_B
    \end{pmatrix}^{-1}\big|_{x_{\rm mid}}
\end{equation}
which relates the periods as $\Pi_A = \mathcal{T}_A^{\; B}\Pi_B $. In turn, one might be interested in another region $C$ that does not overlap with $A$ but does overlap with $B$; for instance, one might have three singularities at $x=0,\mu^{-1},\infty$, and want to go from the vicinity of $x=0$ to $x=\infty$. In that case the transition matrix from $A$ to $C$ is given by first analytically continuing from $A$ to $B$ and then from $B$ to $C$, chaining them together as $\cT_A^{\; C}= \cT_{A}^{\; B}\cT_B^{\;C} $.

\paragraph{Integral basis.} The above two methods allow us to set up local expansions for the periods and determine the transition matrices between all patches. These period vectors, however, are expressed in terms of the Frobenius basis, which generically is a complex rather than an integral basis. In practice one can use the topological data of the mirror Calabi--Yau manifold --- intersection numbers and integrated Chern classes --- to fix the transition matrix $\cT_{\rm LCS}$, transforming us from Frobenius basis to the integral basis at large complex structure, cf.~\eqref{eq:CY4T} and \eqref{eq:CY3T}. This allows one to map the Frobenius periods $\hat{\Pi}_{\rm LCS}$ into the integral basis $\Pi_{\rm LCS}$
\begin{equation}
    \Pi_{\rm LCS} = \cT_{\rm LCS} \hat{\Pi}_{\rm LCS}\, .
\end{equation}
In this work we also consider a complementary approach to obtain this transition matrix for which knowledge of the underlying geometry is not needed. This method uses the monodromy matrices and was employed for instance in \cite{Gerhardus:2016iot}. We use that we already know the integral monodromy matrices of the LCS and conifold point $M_0,M_{\rm C} \in SL(D+1,\bbZ)$. In turn, we can compare these to the monodromies $\hat{M}_0$ and $ \hat{M}_{\rm C}$ in the local Frobenius basis at LCS.\footnote{Note that the monodromy around the conifold point is initially computed in the local Frobenius basis at the conifold point. It should be brought to the Frobenius basis of the LCS point by conjugation by $\hat{\cT}_{\rm C}^{\; \rm LCS}$, which is the transition matrix from the conifold Frobenius basis to the LCS Frobenius basis} The crucial property here is that $M_0,M_{\rm C}$ together uniquely fix the integral frame: the only integral basis transformation $B \in SL(D+1,\bbZ)$ that preserves these monodromies, i.e.
\begin{equation}\label{eq:uniqueB}
B^{-1} M_0 B = M_0 \, , \qquad B^{-1} M_{\rm C} B = M_{\rm C}\, ,
\end{equation}
is given by the identity matrix $B=\mathbb{I}$. For our purposes this means that the transition matrix $\cT_{\rm LCS}$ from the LCS Frobenius basis to the integral basis is fixed uniquely. We refer to the sections \ref{ssec:T2} and \ref{ssec:X6} for two explicit examples, the two-torus $T^2$ and the sextic fourfold. 

\subsection{Warm-up example: \texorpdfstring{$T^2$}{T2}}\label{ssec:T2}
To warm up, we enumerate a particular landscape of elliptic curves and their moduli spaces. The purpose of this section is pedagogical, as we aim to demonstrate our approach for the Calabi--Yau fourfold classification in a simpler setting, and not to give an exhaustive classification of elliptic curves.\footnote{Our assumptions are too strong to reproduce all elliptic curves with three singular fibers. For the original classification we refer to \cite{MirandaPersson}, and a more recent discussion to \cite{Doran:2015xjb}.} In fact, all models we consider here are already known in the physics literature, see for instance \cite{Schimannek:2021pau} for a recent discussion. We have summarized the properties of the four elliptic curves we find in table \ref{table:T2}. For the reader interested in more details on the period computations we also refer to the attached notebook.

\paragraph{Monodromy matrices.} Let us begin by setting up the monodromy matrices $M_i \in SL(2,\bbZ)$ around the three punctures in $\bbP^1 \backslash \{0,\mu^{-1},\infty\}$. Similar to threefold and fourfold case, we want $M_0$ to be maximally unipotent. For $T^2$ this corresponds to a lower-triangular matrix with eigenvalues equal to one, leaving the vector $(0,1)$ invariant. In turn, $M_C$ should shift this invariant state by its magnetic dual $(1,0)$. Altogether this gives us as monodromy matrices
\begin{equation}\label{eq:T2mons}
    M_0 = \begin{pmatrix}
        1 & 0 \\
        1 & 1 
    \end{pmatrix}\, , \qquad M_{\rm C} = \begin{pmatrix}
        1 & -\kappa \\
        0 & 1 
    \end{pmatrix}\, .
\end{equation}
Note that we set the lower-left component of $M_0$ to one, since we want the period vector to transform as $\Pi(t)=(1,t) \to (1,t+1)$ (with $t=\log[z]/2\pi i$) under monodromies. This leaves us with one free parameter $\kappa \in \mathbb{N}$ in $M_{\rm C}$ at our disposal. The monodromy around $z=\infty$ is fixed as the inverse of the product of the first two monodromies as
\begin{equation}
    M_\infty = (M_0 M_{\rm C})^{-1} = \left(
\begin{array}{cc}
 1-\kappa  & \kappa  \\
 -1 & 1 \\
\end{array}
\right)\, ,
\end{equation}
as follows by considering a clockwise loop around $z=\infty$ and interpreting it as a counter-clockwise loop around $z=0$ and $z=\mu^{-1}$.

\paragraph{Quasi-unipotency conditions.} We now investigate for what values of $\kappa \in \mathbb{N}$ the monodromy $M_\infty$ is quasi-unipotent, i.e.~whether it satisfies \eqref{eq:quproperty} for some finite order $l$ and unipotency degree $d$. For the finite orders we need to scan over $l=1,2,3,4,6$, while $d=0,1$ for the unipotency degrees. The conditions that have a solution are then summarized as
\begin{equation}
    \begin{aligned}
         M_\infty^3 - 1 &= (\kappa-3)\left(
\begin{array}{cc}
 2 \kappa -\kappa ^2 & \kappa ^2-\kappa  \\
 1-\kappa  & \kappa  \\
\end{array}
\right)=0\, , \\
         M_\infty^4 - 1 &=(\kappa-2)\left(
\begin{array}{cc}
 \kappa ^3-5 \kappa ^2+5 \kappa  & -\kappa ^3+4 \kappa ^2-2 \kappa  \\
 \kappa ^2-4 \kappa +2 & 3 \kappa -\kappa ^2 \\
\end{array}
\right) = 0\, , \\
         M_\infty^6 - 1 &=(\kappa-1)(\kappa-3)\left(
\begin{array}{cc}
 \kappa ^4-7 \kappa ^3+14 \kappa ^2-7 \kappa  & -\kappa ^4+6 \kappa ^3-9 \kappa ^2+2 \kappa  \\
 \kappa ^3-6 \kappa ^2+9 \kappa -2 & -\kappa ^3+5 \kappa ^2-5 \kappa  \\
\end{array}
\right) = 0\, , \\
         (M_\infty^2 - 1)^2 &=(\kappa-4)\left(
\begin{array}{cc}
 \kappa ^3-3 \kappa ^2+\kappa  & 2 \kappa ^2-\kappa ^3 \\
 \kappa ^2-2 \kappa  & \kappa -\kappa ^2 \\
\end{array}
\right)= 0\, , \\
    \end{aligned}
\end{equation}
which correspond to the values $(l,d)=(3,0),(4,0),(6,0),(2,1)$ respectively. We have extracted all common factors in these conditions, such that the remaining entries do not share any roots for $\kappa$. Taking into account that the solution $\kappa=3$ in the third case actually corresponds to an order $l=3$ monodromy, we find that the respective solutions are $\kappa=3,2,1,4$. 

\begin{table}[t]
\centering
\begin{tabular}{| c | c  c  c  c |  }
\hline & & & & \\[-1.5ex] 
$(a_1,a_2)$  & $(\tfrac{1}{6},\tfrac{5}{6})$ & $(\tfrac{1}{4},\tfrac{3}{4})$ & $(\tfrac{1}{3},\tfrac{2}{3})$ & $(\tfrac{1}{2},\tfrac{1}{2})$  \\[2mm] 
\hline & & & & \\[-1.5ex] 
 $\kappa$ & 1 & 2 & 3 & 4 \\[2mm]
 $\mu$ & 432 & 64 & 27 & 16 \\[2mm]
 $(d,l)$ & $(0,6)$ & $(0,4)$ & $(0,3)$ & $(1,2)$ \\[2mm]
 Modular group & $\Gamma_1(1)$  &  $\Gamma_1(2)$ & $\Gamma_1(3)$  & $\Gamma_1(4)$ \\[2mm]
 Elliptic curve & $X_{6}(1,2,3)$   & $X_{4}(1^2,2)$  & $X_3(1^3)$ & $X_{2,2}(1^4)$ \\[2mm]
 \hline
\end{tabular}
\caption{\label{table:T2} Data for the four elliptic curves with monodromies of the form \eqref{eq:T2mons}. The elliptic curve is indicated by the degrees of the polynomial(s) as subscript and weights of the projective coordinates as argument; for example, $X_{6}(1,2,3)$ is a sextic in $\bbP^{2}[1,2,3]$.}
\end{table}

\paragraph{Picard-Fuchs equation.} Having described the monodromy matrices, we next set up the differential equation for the periods. We want the singularities at $z=0,\mu^{-1}$ both to have local exponents $(0,0)$; this results in one logarithmic period, producing the type of monodromies given in \eqref{eq:T2mons}. On the other hand, the eigenvalues $\lambda_i$ of the monodromy $M_\infty$ fixes the exponents at $x=\infty$ as $2\pi a_i = \arg[\lambda_i]$. Explicitly, we find for the cases $\kappa=1,2,3,4$ respectively that
\begin{equation}
    (a_1,a_2) = \left(\tfrac{1}{6},\tfrac{5}{6}\right), \ \left(\tfrac{1}{4},\tfrac{3}{4}\right), \ \left(\tfrac{1}{3},\tfrac{2}{3}\right), \ \left(\tfrac{1}{2},\tfrac{1}{2}\right)\, .
\end{equation}
Note that the number $\kappa$ may be fixed in terms of the exponents as
\begin{equation}
    \kappa = 4 \sin(2\pi a_1)\sin(2\pi a_2)\, .
\end{equation}
In fact, we find similar relations later for the Calabi--Yau threefolds and fourfolds: all topological data --- the intersection number and Chern classes --- can be determined from the exponents at infinity. We collect the sets of local exponents as a Riemann symbol
\begin{equation}\label{eq:PT2}
{\cal  P} \left\{\begin{array}{ccc}
0& 1/\mu& \infty\\ \hline
0& 0 & a_1\\
0& 0 & a_2\\
\end{array}\right\}\ .
\end{equation}
The differential operator that produces this singularity pattern is given by
\begin{equation}\label{eq:pfT2}
    L = \theta^2-\mu z (\theta+a_1)(\theta+a_2)\,  ,
\end{equation}
with the position of the conifold point also fixed in terms of the exponents as
\begin{equation}\label{eq:T2mu}
    \mu^{-1} = e^{2\gamma_E+\psi(a_1)+\psi(a_2)}\, ,
\end{equation}
with $\gamma_E$ the Euler-Mascheroni constant and $\psi(x)$ the digamma function. 

\paragraph{Frobenius ansatz.} Having given the Picard-Fuchs equation, let us next characterize the Frobenius solutions around the three singularities. Since the singular points at $x=0,\mu^{-1}$ have the same exponents, we can use the same ansatz based on \eqref{eq:frobgen}, which reads
\begin{equation}
\begin{aligned}
    \varpi_0 &= f_0(z)\, , \\
    2\pi i \, \varpi_1 &= f_0(z)\log[z]+f_1(z)\, ,
\end{aligned}
\end{equation}
and where $f_{0,1}(z)$ denote power series
\begin{equation}\label{eq:T2powerseries}
    f_i(z) = \delta_{i,0}+\sum_{k=1}^\infty c_{i,k} z^k\, ,
\end{equation}
with $c_{i,k}$ rational coefficients. For $z=\infty$ it depends on whether $a_1=a_2$ or not. When $a_1=a_2=\tfrac{1}{2}$ the solution is similar to the points  $z=0,\mu^{-1}$, differing by a factor of $\sqrt{z}$, namely
\begin{equation}
\begin{aligned}
    \varpi_0 &= \sqrt{z} f_0(z)\, , \\
    2\pi i \, \varpi_1 &= \sqrt{z}f_0(z)\log[z]+\sqrt{z}f_1(z)\, ,
\end{aligned}
\end{equation}
where $f_{0,1}$ has a similar series expansions \eqref{eq:T2powerseries}. For $a_1\neq a_2$ we build two different solutions
\begin{equation}
\begin{aligned}
    \varpi_0 &= z^{a_1} f_0(z)\, , \\
    \varpi_1 &= z^{a_2}f_1(z)\, ,
\end{aligned}
\end{equation}
where in both cases the power series is given by an expansion of the form
\begin{equation}
    f_i(z) = 1+\sum_{k=1}^\infty c_{i,k} z^k\, .
\end{equation}
By plugging these series ans\"atze into the Picard-Fuchs equation of the four cases, we can solve order-by-order for the coefficients.

\paragraph{Hypergeometric solution.} We can also solve the Picard-Fuchs equation in terms of hypergeometric functions. This gives an instructive look into the series coefficients of the Frobenius solution around $z=0$; see also \cite{Zagier} for a discussion on the modularity underlying these series, labelled as A, B, C and D. In general, we can write the solutions to \eqref{eq:pfT2} as
\begin{equation}
    \varpi_0 = {}_2 F_1\left(a_1,a_2;1;\mu  z\right)\, , \qquad \varpi_1 = \frac{i}{\sqrt{\kappa}} \cdot {}_2 F_1\left(a_1,a_2;1;1-\mu  z\right)\, .
\end{equation}
The fundamental period $\varpi_0$ admits an expansion as infinite series
\begin{equation}
\begin{aligned}
    \kappa=1&: \qquad \varpi_0  = \sum_{n=0}^\infty \frac{(6n)!}{n!(2n)!(3n)!} z^n  = 1+60 z + 13860 z^2+4084080 z^3 + \mathcal{O}(z^4) \, , \\
    \kappa=2&: \qquad \varpi_0  = \sum_{n=0}^\infty \frac{(4n)!}{(n!)^2(2n)!} z^n = 1+12 z + 420 z^2+ 18480 z^3 + \mathcal{O}(z^3) \, , \\
    \kappa=3&: \qquad \varpi_0  = \sum_{n=0}^\infty \frac{(3n)!}{(n!)^3} z^n  = 1 + 6 z + 90 z^2+ 1680 z^3 + \mathcal{O}(z^4)\, , \\
    \kappa=4&: \qquad \varpi_0  = \sum_{n=0}^\infty \frac{((2n)!)^2}{(n!)^4} z^n  = 1+ 4z+36 z^2+400 z^3 +\mathcal{O}(z^4)\, .
\end{aligned}
\end{equation} 
From these series we may also read off the corresponding complete intersections in weighted projective space, as explained in section \ref{ssec:CICYs}. The numerators specify the degrees and the denominators the weights of the projective spaces, giving us the geometries listed in table \ref{table:T2}. For the logarithmic period $\varpi_1$ we can obtain similar series expansions
\begin{equation}
\begin{aligned}
    \kappa=1&: \qquad \varpi_1  = \frac{1}{2\pi i}\left( \varpi_0 \log[z]+312 z + 77652 z^2 + 23485136 z^3 +\mathcal{O}(z^4)\right)  \, , \\
    \kappa=2&: \qquad \varpi_1  = \frac{1}{2\pi i}\left( \varpi_0 \log[z]+ 40 z + 1556 z^2 + \frac{213232}{3} z^3+\mathcal{O}(z^4)\right)  \, , \\
    \kappa=3&: \qquad \varpi_1  = \frac{1}{2\pi i}\left( \varpi_0 \log[z]+15 z + \frac{513}{2} z^2 + 5018 z^3 +\mathcal{O}(z^4)\right)  \, , \\
    \kappa=4&: \qquad \varpi_1  = \frac{1}{2\pi i}\left( \varpi_0 \log[z]+ 8 z + 84 z^2 + \frac{2960}{3} z^3+\mathcal{O}(z^4)\right)  \, .
\end{aligned}
\end{equation}

\paragraph{Transition matrix.} Having set up the local period solutions, let us next discuss the transition matrices between different patches. We focus only at the transition matrix between $z=0$ and $z=\mu^{-1}$. We find that we can give the result for all four values of $\kappa$ in the succinct form
\begin{equation}
    \cT_{0}^{\; \rm C} = \frac{i}{2\sqrt{\kappa}}\begin{pmatrix}
        \kappa & -2\kappa \\
        2 & 0
    \end{pmatrix}\, ,
\end{equation}
which we computed following the methodology laid out in section \ref{ssec:localperiods}. We can check straightforwardly how this transforms the Frobenius monodromy around $z=\mu^{-1}$ into the Frobenius basis of the LCS point. In the Frobenius basis the conifold monodromy is simply lower-triangular with a one in the lower-left corner, which transforms under the transition matrix as
\begin{equation}
    M_{\rm C} =     \cT_{0}^{\; \rm C} \begin{pmatrix}
        1 & 0 \\
        1 & 1 
    \end{pmatrix}(    \cT_{0}^{\; \rm C} )^{-1} = \begin{pmatrix}
        1 & -\kappa \\
        0 & 1
    \end{pmatrix}\, .
\end{equation}
Notice that this is precisely the integral monodromy we started from in \eqref{eq:T2mons}. Since the monodromy in the Frobenius basis around $z=0$ also agrees with \eqref{eq:T2mons}, we are left to conclude that the LCS Frobenius basis coincides with the integral basis. Let us stress that this is special to these elliptic curves, and not true for the Calabi--Yau threefolds and fourfolds studied later.

\paragraph{Modular curve.} The complex structure moduli spaces given here have a natural interpretation in terms of modular curves $\bbH/\Gamma$, where $\bbH$ denotes the upper-half plane. The monodromy group $\Gamma \subseteq SL(2,\bbZ)$ is generated by the monodromies given in \eqref{eq:T2mons}, which yields 
\begin{equation}
    \Gamma_1(\kappa) = \left\{ \begin{pmatrix}
        a & b \\
        c & d 
    \end{pmatrix} \in SL(2,\bbZ)\ | \  \begin{pmatrix}
        a & b \\
        c & d
    \end{pmatrix} = \begin{pmatrix}
        1 & 0 \\
        \ast & 1
    \end{pmatrix} \mod \kappa \right\}\, .
\end{equation}
The constraint on the upper-right entry follows directly from the fact that $M_C$ can only shift it by units of $\kappa$. The resulting modular curves are commonly denoted by $X_1(\kappa) = \bbH/\Gamma_1(\kappa)$. For $\kappa=1$ we have $\Gamma_1(1)=\rm{SL}(2,\bbZ)$, so the standard fundamental domain $X_1(1)=\bbH/\rm{SL}(2,\bbZ)$.

\section{Classification of Calabi--Yau fourfolds}\label{sec:classification}
In this section we enumerate all Calabi--Yau fourfolds $Y_4$ whose complex structure moduli space is the thrice-punctured sphere
\begin{equation}
    \cM_{\rm cs}(Y_4)= \bbP^1 \backslash \{0,\mu^{-1},\infty\}\, ,
\end{equation}
and subject to the additional requirement that  $z=0$ and $z=\mu^{-1}$ are large complex structure and conifold points. In section \ref{ssec:classmonodromies} we characterize the monodromy matrices and classify all possible monodromy tuples of that type, finding 14 possibilities as summarized in table \ref{table:monodromydata}. The next subsection \ref{ssec:CY4localperiods} studies the periods associated to these monodromies and extracts the topological data from our integral basis. In section \ref{ssec:CICYs} we identify the 14 mirror Calabi--Yau fourfolds as complete intersections in weighted projective spaces, and also point out relations to the 14 hypergeometric Calabi--Yau threefolds. In the final subsection \ref{ssec:X6} we discuss the mirror sextic fourfold as a working example. We have attached two notebooks for the reader interested in the classification and the computation of the periods.

\subsection{Classification of monodromy matrices}\label{ssec:classmonodromies}
We begin with the classification of the monodromy tuples. We first set up the monodromy matrices around the large complex structure and conifold point, parametrized by two integer parameters $\kappa,a$. We then invoke that the monodromy matrix for the singular point $z=\infty$ must be quasi-unipotent \eqref{eq:quproperty}, resulting in fourteen possible values for $\kappa, a$. The results of this classification are summarized in table \ref{table:monodromydata}. We refer to appendix \ref{app:Ktheory} for the match of our basis to the K-theory basis of \cite{Gerhardus:2016iot}.

\paragraph{Hodge numbers.} We begin by fixing the Hodge numbers in our variation of Hodge structure. We consider Calabi--Yau fourfolds $Y_4$ that have a primitive middle cohomology $H^4_{\rm prim}(Y_4, \bbZ)$ with Hodge numbers 
\begin{equation}\label{eq:hpq}
    (h^{4,0}, h^{3,1}, h^{2,2}, h^{1,3}, h^{0,4})  = (1,1,1,1,1)\, .
\end{equation}
In principle one could consider a larger $h^{2,2}>1$, but for our purposes we can ignore these additional $(2,2)$-forms:  the types of monodromies we consider would leave these states invariant, and so it follows from the rigidity theorem of \cite{Schmid} that we can project out them out (analogous to non-primitive $(2,2)$-forms). 

\paragraph{Intersection pairing.} Let us next write down the intersection pairing $\eta$. From these dimensions in \eqref{eq:hpq} we read off $(3,2)$ as signature of the pairing, since $(4,0)$-forms and $(2,2)$-forms have positive pairing with their conjugates, while for $(3,1)$-forms this pairing is negative. We write this pairing of SO$(3,2;\bbZ)$ as
\begin{equation}\label{eq:eta}
    \eta = \left(
\begin{array}{ccccc}
 0 & 0 & 0 & 0 & 1 \\
 0 & 0 & 0 & 1 & 0 \\
 0 & 0 & \kappa  & 0 & 0 \\
 0 & 1 & 0 & 0 & 0 \\
 1 & 0 & 0 & 0 & 0 \\
\end{array}
\right)\, .
\end{equation}
It is parametrized by a positive integer $\kappa \in \bbN$, which from a geometrical perspective corresponds to the intersection number of the mirror Calabi--Yau fourfold. Monodromy matrices $M_i$ satisfy
\begin{equation}\label{eq:etacompatible}
    (M_i)^T \eta M_i = \eta\, ,
\end{equation}
as they lie in the isometry group of the pairing $M_i \in SO(3,2; \bbZ)$.

\paragraph{Large complex structure monodromy.} We begin with the monodromy $M_0 \in SO(3,2;\bbZ)$ around the large complex structure point. This should be a matrix that is maximally unipotent and all eigenvalues equal to one, where the former means that the log-monodromy matrix $N_0 = \log M_0$ is maximally nilpotent, i.e.~$(N_0)^4 \neq 0$ and $(N_0)^5=0$. By pairing-preserving transformations we can always bring such a matrix to a lower-triangular form with $1$'s on the diagonal.
In turn, imposing compatibility with the pairing \eqref{eq:etacompatible} yields a monodromy matrix
\begin{equation}
    M_0 = \left(
\begin{array}{ccccc}
 1 & 0 & 0 & 0 & 0 \\
 -d & 1 & 0 & 0 & 0 \\
 b-c d & c  & 1 & 0 & 0 \\
 \frac{\kappa}{2} d c^2  +b c \kappa +a & -\frac{c^2 \kappa }{2} & -c\kappa & 1 & 0 \\
 a d-\frac{ \kappa }{2}b^2 & -a & -b \kappa & d & 1 \\
\end{array}
\right)\, , 
\end{equation}
with four integral parameters $a,b,c,d \in \bbZ$. We simplify this monodromy matrix by setting two parameters to one: $c=1$ and $d=1$. We then consider the basis transformation
\begin{equation}\label{eq:B}
    B = \left(
\begin{array}{ccccc}
 1 & 0 & 0 & 0 & 0 \\
 0 & 1 & 0 & 0 & 0 \\
 0 & e   & 1 & 0 & 0 \\
 0 & -\frac{e^2 \kappa }{2} & -e \kappa & 1 & 0 \\
 0 & 0 & 0 & 0 & 1 \\
\end{array}
\right) \in SO(3,2;\bbZ)\, : \qquad M_0 \to B^{-1} M_0 B\, ,
\end{equation}
which can be interpreted as shifts $b\to b+e$ and $a \to a +\tfrac{1}{2}e^2\kappa$. We choose $e=-b$ such that we set $b=0$, and then redefine $a \to a+\tfrac{1}{2}b^2 \kappa$. Altogether this yields as  monodromy
\begin{equation}\label{eq:M0}
    M_0 = \left(
\begin{array}{ccccc}
 1 & 0 & 0 & 0 & 0 \\
 -1 & 1 & 0 & 0 & 0 \\
 -1  & 1 & 1 & 0 & 0 \\
 a+\frac{\kappa }{2} & -\frac{\kappa }{2} & -\kappa & 1 & 0 \\
 a & -a & 0 & 1 & 1 \\
\end{array}
\right)\, ,
\end{equation}
with integer parameter $a \in \bbZ$. Integrality of $M_0 \in SO(3,2;\bbZ)$ demands $\frac{\kappa}{2} \in \bbN$ as a quantization condition, implying that $\kappa$ has to be even. We also want to point out that this lower-triangular form of $M_0$ has a natural interpretation from the dual perspective of Type IIA compactified on the mirror Calabi--Yau fourfold, where the basis vectors used here correspond to the D0, D2, D4, D6, and D8-brane charges. 

\paragraph{Conifold monodromy.} We now consider the monodromy around the conifold point. In contrast to Calabi--Yau threefolds, its monodromy is of order two instead of infinite \cite{Grimm:2009ef}. What this monodromy does is swapping the mirror D0 and D8-branes. IN the mathematics literature this monodromy referred to as a Seidel-Thomas twist \cite{Seidel:2000ia}, and a recent discussion in the physics context can be found in \cite{Gerhardus:2016iot}. We follow the basis convention used in \cite{CaboBizet:2014ovf}, for which $M_C$ reads
\begin{equation}\label{eq:MC}
    M_{C} = \left(
\begin{array}{ccccc}
 0 & 0 & 0 & 0 & -1 \\
 0 & 1 & 0 & 0 & 0 \\
 0 & 0 & 1 & 0 & 0 \\
 0 & 0 & 0 & 1 & 0 \\
 -1 & 0 & 0 & 0 & 0 \\
\end{array}
\right)\, .
\end{equation}
From a physical perspective, the corresponding light state is given by the charge vector $q=(1,0,0,0,1)$. This can be read off from  \eqref{eq:MC} as this state lies in the image of $N_{1}  = 1-M_{1}$. We also want to note that $M_C$ is left invariant by the basis transformation \eqref{eq:B}. In particular, our choice of setting $b=0$ thus did not affect the choice of conifold monodromy.

\paragraph{Other integral lattices.} Let us briefly comment on extending the basis set up here to other inequivalent integral lattices. These other lattices may be obtained by considering non-integral basis transformations in GL$(5,\bbR)$ that keep the monodromy matrices \eqref{eq:M0}, \eqref{eq:MC} and pairing \eqref{eq:eta} integral. Such transformations oftentimes happen when we consider orbifolds of Calabi--Yau manifolds. An example is the mirror quintic threefold and its $\bbZ_5$-quotiented twin, which have different integral lattices for the middle cohomology, but the same functional dependence for their periods. At the level of the monodromy this corresponds to $\kappa,a$ sharing a common divisor, and in turn applying a basis transformation that rescales the corresponding entries, possibly setting $c,d\neq 1$. In \cite{Doran:2005gu} this was readily worked out in the threefold case, but we leave its implementation to fourfolds for future work. We note, however, that we already encountered an example that goes beyond the ansatz  \eqref{eq:M0}, as for the example with two LCS points we found a different form for the monodromy \eqref{eq:MLCS}.

\paragraph{Monodromy at infinity.} With these two monodromies fixed, the monodromy around $z=\infty$ follows by considering an anti-clockwise oriented loop around $z=0$ and $z=1$, since we are dealing with the thrice-punctured sphere $\bbP^1\backslash \{0,1,\infty\}$ as moduli space. Thus $M_\infty$ is computed by the inverse of the product of the first two monodromies
\begin{equation}\label{eq:Minfty}
    M_\infty = (M_{\rm 0} M_{1})^{-1} = \left(
\begin{array}{ccccc}
 -a & -a-\frac{\kappa }{2} & \kappa & 1 & -1 \\
 1 & 1 & 0 & 0 & 0 \\
 0 & -1  & 1 & 0 & 0 \\
 -a & -\frac{\kappa }{2} & \kappa & 1 & 0 \\
 -1 & 0 & 0 & 0 & 0 \\
\end{array}
\right)\, .
\end{equation}
In order for this $M_\infty$ to be a well-defined monodromy matrix in SO$(3,2;\bbZ)$, it must be a quasi-unipotent matrix. Let us briefly recall this condition \eqref{eq:quproperty}, given by
\begin{equation}
    (M_\infty^l-1)^d \neq 0\, , \qquad (M_\infty^l-1)^{d+1}=0\, .
\end{equation}
Our setting fixes what indices $l,d$ are allowed: the fact that we are dealing with Calabi--Yau fourfolds constrains the nilpotency degree $d$, while working with rational matrices restricts the orders $l$. The values that need to be considered are
\begin{equation}
\begin{aligned}
    d=0&: \qquad l=1,2,3,4,5,6,8,10,12\,, \\
    d=1,2,3&: \qquad l=1,2,3,4,5,6\,, \\
    d=4&: \qquad l=1,2\,. \\
\end{aligned}
\end{equation}
The values of $l$ accompanying the different values of $d$ follow from the fact that the unipotent part $M_u$ and the semisimple part $M_{ss}$ of the monodromy need to commute.\footnote{The reason is that when $M_u$ is non-trivial, we can decompose our vector space into irreducible representations under $M_u$, and the finite order matrix $M_{ss}$ has to act with the same eigenvalue on all elements in such a representation. This is related to the fact that the semisimple monodromy has to be an automorphism of the limiting mixed Hodge structure, see appendix \ref{app:lmhs}.}  
As detailed in the accompanying notebook in the construction, we scan over these values of $l,d$ and solve for pairs $(\kappa,a)$ in each case. This results in 14 solutions summarized in table \ref{table:monodromydata}. We find four singularity types at infinity: 7 cases that have finite order monodromies $d=0$ and $l=6,8,10,12$, 3 with nilpotency degree $d=1$ and orders $l=4,6$, 3 with $d=2$ and $l=4,6$, and finally one case with $d=4$ and order $l=2$. Note that this is the same partition as arises for the fourteen hypergeometric Calabi--Yau threefolds \cite{Doran:2005gu, almkvist2005tables}, based on whether there is a finite monodromy point, K-point, conifold point or another LCS point at infinity.

\begin{table}[t]
\centering
\renewcommand*{\arraystretch}{1.5}
\begin{subtable}{\textwidth}
\centering
\begin{tabular}{| c || c | c | c | c | c | c | c | c | }
\hline  $(\kappa,a)$ & (6,4) & (4,4) & (2,3) & (10,5) & (2,4) & (4,3) & (12,5)  \\ \hline
degree $d$ & \multicolumn{7}{|c|}{0} \\ \hline
order $l$ & 6 & 8 & \multicolumn{2}{|c|}{10} & \multicolumn{3}{|c|}{12} \\ \hline
\end{tabular}
\subcaption{\label{table:finite}Finite order monodromies.}
\end{subtable}
\begin{subtable}{\textwidth}
\centering
\begin{tabular}{| c || c | c | c | c | c | c | c | c | }
\hline $(\kappa,a)$ & (8,4) & (2,2) & (18,6) & (16,6) & (8,5) & (24,7) & (32,8) \\ \hline
degree $d$ & \multicolumn{3}{|c|}{1} & \multicolumn{3}{|c|}{2} & 4 \\ \hline
order $l$ & 4 & \multicolumn{2}{|c|}{6} & 4 & \multicolumn{2}{|c|}{6} & 2 \\ \hline
\end{tabular}
\subcaption{\label{table:infinite}Infinite order monodromies.}
\end{subtable}
\caption{\label{table:monodromydata}Monodromy data for Calabi-Yau fourfolds with $\cM_{\rm cs}=\bbP^1\backslash\{0,\mu^{-1},\infty\}$. In \ref{table:finite} we collected all cases with a finite order $M_\infty$ while \ref{table:infinite} contains all infinite order monodromies. The integers $(\kappa,a)$ specify the monodromy around large complex structure \eqref{eq:M0}, while $(d,l)$ denote the quasi-unipotency indices in \eqref{eq:quproperty} for $M_\infty$. Geometrically $\kappa$ is the mirror intersection number, while $a=(\kappa+c_2)/24$ is related to the second Chern class.}
\end{table}

\subsection{Local periods and transition matrices}\label{ssec:CY4localperiods}
Having set up the monodromy matrices, we now study the periods giving rise to these monodromy groups. We give the Picard-Fuchs equations, set up the local Frobenius solutions, and explain how to fix the integral basis from the monodromies. For the interested reader we refer to the accompanying notebook on fourfold periods, where one can simply select a model and execute the notebook to compute the periods and K\"ahler potential in all three phases.

\paragraph{Picard-Fuchs equation.}
The periods of the Calabi--Yau fourfolds considered in this work satisfy differential equations of order five.\footnote{For some Calabi--Yau fourfolds they are of order six, corresponding to manifolds whose LCS monodromy is not maximally unipotent. We refer to \cite{Gerhardus:2016iot} for a detailed study of such examples.} In the literature these sorts of differential operators for periods are commonly referred to as Picard-Fuchs operators. For the cases at hand they have three singularities. The local exponents of the solutions at $z=0$ and $z=\mu^{-1}$ are fixed to be $(0,0,0,0,0)$ and $(0,1, 2,3, \tfrac{3}{2})$. The exponents $a_i$ at the remaining singularity at infinity may be computed from the eigenvalues $\lambda_i$ of the monodromy at infinity $M_\infty$ as
\begin{equation}\label{eq:expsdef}
    a_i = \frac{1}{2\pi}\arg[\lambda_i] \, ,
\end{equation}
where we take $0 \leq a_i < 1$. We choose to sort the $(a_1,a_2,a_3,a_4,a_5)$ in ascending order. We find that $a_3=\tfrac{1}{2}$ in all cases, and $a_4=1-a_1$ and $a_5=1-a_2$. Taking the matrices in the same order as given in table \ref{table:monodromydata}, we find the independent exponents to respectively be
\begin{equation}
\begin{aligned}
    (a_1,a_2) = &\Big(\frac{1}{6},\frac{1}{3}\Big)\, , \ \Big(\frac{1}{8}, \frac{3}{8}\Big)\, , \ \Big(\frac{1}{10}, \frac{3}{10}\Big)\, , \ \Big(\frac{1}{5}, \frac{2}{5}\Big)\, , \ \Big(\frac{1}{12}, \frac{5}{12}\Big)\, , \ \Big(\frac{1}{6}, \frac{1}{4}\Big)\, , \ \Big(\frac{1}{4}, \frac{1}{3}\Big)\, , \\
    &\Big(\frac{1}{4},\frac{1}{4}\Big)\, , \ \Big(\frac{1}{6}, \frac{1}{6}\Big)\, , \ \Big(\frac{1}{3}, \frac{1}{3}\Big)\, , \ \Big(\frac{1}{4}, \frac{1}{2}\Big)\, , \ \Big(\frac{1}{6}, \frac{1}{2}\Big)\, , \ \Big(\frac{1}{3}, \frac{1}{2}\Big)\, , \ \Big(\frac{1}{2}, \frac{1}{2}\Big)\, , \\
\end{aligned}
\end{equation}
For every example we can collect the exponents of the three singularities into a Riemann symbol
\begin{equation}\label{eq:PCY4}
{\cal  P} \left\{\begin{array}{ccc}
0& 1/\mu& \infty\\ \hline
0& 0 & a_1\\
0& 1 & a_2\\
0& 2 & a_3\\
0& 3 & a_4\\
0& 3/2 & a_5
\end{array}\right\}\ .
\end{equation}
Given the local exponents, we can write down the differential operator for the periods. It may be given in terms of the exponents $a_i$ at infinity as\begin{equation}\label{eq:PFfourfold}
    L = \theta^5 - \mu z (\theta+a_1)\cdots (\theta+a_5)\, ,
\end{equation}
By rescalings of $z$ we may set the singularity at $z=\mu^{-1}$ to the convenient position
\begin{equation}\label{eq:mu}
    \mu^{-1} = e^{5\gamma_E+\psi(a_1)+\psi(a_2)+\psi(a_3)+\psi(a_4)+\psi(a_5)}\, ,
\end{equation}
where $\gamma_E$ denotes the Euler-Mascheroni constant and $\psi(x)$ the digamma function. This choice removes terms involving $\log \mu$ from the periods and transition matrices.

\paragraph{Frobenius solutions.} The series expansion of the periods around the singularities are known as Frobenius solutions. Based on \eqref{eq:frobgen} the expansion around the LCS point reads
\begin{align}\label{eq:frobLCS}
    \varpi_0 &= f_0(z)\, ,\nonumber \\
    2\pi i\, \varpi_1 &= f_0(z) \log[z]+f_1(z)\, , \nonumber \\
    (2\pi i)^2\,\varpi_2 &= f_0(z) \log[z]^2 + 2 f_1(z) \log[z]+ f_2(z) \, , \nonumber \\
    (2\pi i)^3\,\varpi_3 &= f_0(z) \log[z]^3 + 3 f_1(z) \log[z]^2+ 3 f_2(z) \log[z]+f_3(z)\, , \\
    (2\pi i)^4\,\varpi_4 &= f_0(z) \log[z]^4+ 4 f_1(z) \log[z]^3+ 6 f_2(z) \log[z]^2 +4 f_3(z) \log[z]+f_4(z)\, ,\nonumber 
\end{align}
where we defined the holomorphic power series
\begin{equation}\label{eq:seriesLCS}
    f_i(z) = \delta_{i,0}+\sum_{k=1}^\infty c_{i,k}z^k\, .
\end{equation}
By plugging these power series into the Picard-Fuchs equation, one can solve for the coefficients $c_{i,k}$ order-by-order.
We can perform a similar expansion around the conifold point $z=\mu^{-1}$. Shifting $z\to z+\mu^{-1}$ such that it is located at $z=0$, we write the series expansions
\begin{equation}\label{eq:frobC}
\begin{aligned}
       \varpi_0 &=   f_0(z)\, , \quad  &\varpi_1 &=   z f_1(z)\, , \quad &\varpi_{\frac{3}{2}} &=   z^{\frac{3}{2}} f_{\frac{3}{2}}(z)\, , \quad &\varpi_2 &=   z^2f_2 (z)\, , \quad &\varpi_3 &=   z^3f_3 (z) \, ,\\
\end{aligned}
\end{equation}
where all power series are given by
\begin{equation}
f_a(z) = 1+\sum_{k=1}^\infty c_{a,k}z^k\, , \qquad a=0,1,2,3,\tfrac{3}{2}\, .
\end{equation}
For the integer exponent $a=0,1,2,3$ we may furthermore set the coefficients $c_{i,k}$ with $i+k\leq 3$ to zero, such that the first non-trivial coefficient in $\varpi_0, \ldots, \varpi_3$ appears at order $z^4$. We can write the monodromy matrices in the Frobenius basis for these two singularities as
\begin{equation}\label{eq:Mfrob}
    \hat{M}_0 = \begin{pmatrix}
        1 & 0 & 0 & 0 & 0 \\
        1 & 1 & 0 & 0 & 0 \\
        1 & 2 & 1 & 0 & 0 \\
        1 & 3 & 3 & 1 & 0 \\
        1 & 4 & 6 & 4 & 1 \\
    \end{pmatrix}\, , \qquad \hat{M}_{C} = \begin{pmatrix}
        1 & 0 & 0 & 0 & 0 \\
        0 & 1 & 0 & 0 & 0 \\
        0 & 0 & -1 & 0 & 0 \\
        0 & 0 & 0 & 1 & 0 \\
        0 & 0 & 0 & 0 & 1 \\
    \end{pmatrix}\, .
\end{equation}
Finally, there is the singular point at infinity. Here we can have four different singularity types: 
\begin{itemize}
    \item F-point: a point with finite order monodromy, also known as a Landau-Ginzburg point sometimes. These points have distinct exponents $a_1\neq a_2\neq a_3\neq a_4\neq a_5$, such that this coordinate singularity may be resolved by going to a finite cover.
    \item C-point: a finite distance point with unipotent monodromy factor of degree $k=3$. The middle three exponents are equal $a_1 \neq a_2=a_3=a_4\neq a_5$. This terminology stems from the fact that C-points resemble the conifold points of Calabi--Yau threefolds.
    \item CY3-point: an infinite distance point with unipotent monodromy factor of degree $k=2$. The outer two exponents are equal  $a_1 = a_2\neq a_3\neq a_4 = a_5$. The term CY3 stands for the fact that the limiting mixed Hodge structure contains two rows with Hodge numbers $(1,0,0,1)$, resembling the middle cohomology of a rigid Calabi--Yau threefold.
    \item LCS-point: an infinite distance point with maximally unipotent monodromy, i.e.~another large complex structure point. All exponents are equal $a_1 = a_2= a_3= a_4 = a_5$.
\end{itemize}

\paragraph{Transition matrix.} Here we apply the methodology of section \ref{ssec:localperiods} to determine the transition matrix to the integral basis. We went through all 14 cases in table \ref{table:data} and observed that the transition matrix from the Frobenius basis at LCS to the integral basis is given by
\begin{equation}\label{eq:CY4T}
    \mathcal{T}_{\rm LCS} =\left(
\begin{array}{ccccc}
 1 & 0 & 0 & 0 & 0 \\
 0 & -1 & 0 & 0 & 0 \\
 \frac{c_2}{24 \kappa } & -\frac{1}{2} & -\frac{1}{2} & 0 & 0 \\
 -\frac{c_2}{48}+\frac{i c_3 \zeta (3)}{8 \pi ^3} & \frac{\kappa }{8} & \frac{\kappa }{4} & \frac{\kappa }{6} &
   0 \\
 -\frac{c_4}{3456}-\frac{5}{12} & \frac{i c_3 \zeta (3)}{8 \pi ^3} & \frac{c_2}{48} & 0 & \frac{\kappa }{24} \\
\end{array}
\right)\, .
\end{equation}
All numbers in this transition matrix follow from the exponents $a_1,\ldots,a_5$ in the differential operator: the number $\kappa$ is computed straightforwardly through sines as
\begin{equation}\label{eq:ksine}
    \kappa = 2^5 \prod_{i=1}^5\sin(a_i)\, ,
\end{equation}
while $c_2$ and $c_3$ may be computed from the formulas
\begin{equation}\label{eq:c2c3}
    c_2 = \kappa \left(-\frac{5}{2} +\frac{3}{\pi^2} \sum_i \psi_1(a_i)\right)\, , \qquad c_3 = -\frac{\kappa}{6}\left( 10 + \frac{1}{\zeta(3)} \sum_i \psi_2(a_i) \right)\, .
\end{equation}
where $\psi_i$ denotes the $i$-th polygamma function. All these relations can be explained from the fact that the periods may be expressed as hypergeometric functions, with these numbers simply showing up in their series expansions. In comparison to the parametrization of the monodromy matrices before we note that $a=(\kappa+c_2)/24$.

\begin{table}[!t]
\centering
\begin{tabular}{| c | c | c | c | c | c | c | c | }
\hline   $a_1,a_2,a_3,a_4,a_5$ & Type & Mirror & $\mu$ & $(\kappa,a)$ & $c_2$ & $c_3$ & $c_4$ 
\\
\hline & & & & & & &  \\[-1.5ex]
$\frac{1}{5}, \frac{2}{5}, \frac{1}{2}, \frac{3}{5}, \frac{4}{5}$ & F & $X_{2,5}(1^7)$ & $2^2 5^5$ &$(10,5)$  & 110 & $-420$ & 2190 
  \\[2mm]
$\frac{1}{10}, \frac{3}{10}, \frac{1}{2}, \frac{7}{10}, \frac{9}{10}$ & F &  $X_{10}(1^5,5)$ & $2^{10} 5^5$ &$(2,3)$ & 70 & $-580$ & 5910 
\\[2mm]
 $\frac{1}{2}, \frac{1}{2}, \frac{1}{2}, \frac{1}{2}, \frac{1}{2}$ & LCS &  $X_{2^5}(1^{10})$ & $2^{10}$ &$(32,8)$ & 160 & $-320$ & 960 
 \\[2mm]
$\frac{1}{3}, \frac{1}{3}, \frac{1}{2}, \frac{2}{3}, \frac{2}{3}$ & CY3 & $X_{2,3,3}(1^8)$ & $2^2 3^6$ & $(18,6)$ & 126 & $-324$ & 1206 
\\[2mm]
$\frac{1}{3}, \frac{1}{2}, \frac{1}{2}, \frac{1}{2}, \frac{2}{3}$ & C & $X_{2,2,2,3}(1^9)$ & $2^6 3^3$ &$(24,7)$ & 144 & $-336$ & 1152 
 \\[2mm]
$\frac{1}{4}, \frac{1}{2}, \frac{1}{2}, \frac{1}{2}, \frac{3}{4}$ & C &  $X_{2,2,4}(1^8)$ & $2^{12}$ &$(16,6)$ & 128 & $-384$ & 1632 
\\[2mm]
 $\frac{1}{8}, \frac{3}{8}, \frac{1}{2}, \frac{5}{8}, \frac{7}{8}$ & F & $X_{2,8}(1^6,4)^*$ &  $2^{18}$ &$(4,4)$ & 92 & $-600$ & 4908 
 \\[2mm]
$\frac{1}{6}, \frac{1}{3}, \frac{1}{2},\frac{2}{3},\frac{5}{6}$ & F & $X_6(1^6)$ &  $6^6$ &$(6,4)$  & 90 & $-420$ & 2610 
\\[2mm]
 $\frac{1}{12}, \frac{5}{12}, \frac{1}{2}, \frac{7}{12}, \frac{11}{12}$ & F & $X_{2,2,12}(1^6,4,6)^{**}$ & $2^{14} 3^6$ &$(2,4)$ &   94 & $-972$ & 11814 
 \\[2mm]
$\frac{1}{4}, \frac{1}{4}, \frac{1}{2}, \frac{3}{4}, \frac{3}{4}$ & CY3 & $X_{4,4}(1^6,2)$ & $2^{14}$  & $(8,4)$ & 88 & $-304$ & 1464 
 \\[2mm]
 $\frac{1}{4}, \frac{1}{3}, \frac{1}{2}, \frac{2}{3}, \frac{3}{4}$ & F & $X_{3,4}(1^7)$  & $2^8 3^3$ &$(12,5)$ & 108 & $-336$ & 1476 
 \\[2mm]
 $\frac{1}{6}, \frac{1}{4}, \frac{1}{2}, \frac{3}{4}, \frac{5}{6}$ & F & $X_{4,6}(1^5,2,3)^*$&  $2^{12} 3^3$ &$(4,3)$  & 68 & $-320$ & 2028 
\\[2mm] 
 $\frac{1}{6}, \frac{1}{6}, \frac{1}{2}, \frac{5}{6}, \frac{5}{6}$  & CY3 &$X_{6,6}(1^4,2,3^2)^*$  & $2^{10}3^3$ &$(2,2)$ & 46 & $-244$ & 1734 
 \\[2mm] 
$\frac{1}{6}, \frac{1}{2}, \frac{1}{2}, \frac{1}{2}, \frac{5}{6}$ & C & $X_{2,2,6}(1^7,3)^*$ & $2^{10}3^6$ &$(8,5)$ &  112 & $-528$ & 3264 
\\[2mm]

 \hline
\end{tabular}
\caption{\label{table:data}All Calabi--Yau fourfolds with $\cM_{\rm cs}=\bbP^1-\{0,\mu^{-1},\infty\}$ and standard LCS and conifold points. The new mirror Calabi--Yau fourfolds compared to \cite{CaboBizet:2014ovf} are indicated by $^*$, with one new geometry being singular and indicated by $^{**}$. }
\end{table}

\paragraph{Topological data.} The above model-dependent data can be related to the topological data of the mirror Calabi--Yau fourfold $X_4$. The intersection number is defined as
\begin{equation}
    \kappa  = \int_{X_4} D \wedge D \wedge D \wedge D \, , 
\end{equation}
where $D$ is the $(1,1)$-form dual to the divisor class generating the K\"ahler cone. Topological numbers related to the Chern classes are obtained as
\begin{equation}\label{eq:chernclasses}
    c_2 = \int_{X_4} c_2(X_4) \wedge D \wedge D\, , \qquad c_3 =\int_{X_4} c_3(X_4) \wedge D \, , \qquad
    c_4 = \int_{X_4} c_4(X_4)\, .
\end{equation}
We note that only $\kappa,c_2,c_3$ are independent, as all other integrals may be expressed in terms of these three numbers: for the integral of the square of the second Chern class we have
\begin{equation}
    \int_{X_4} c_2(X_4)^2= \frac{c_2^2}{\kappa}\, .
\end{equation}
On the other hand, the number $c_4$ follows from the identity \cite{Grimm:2009ef, CaboBizet:2014ovf}\footnote{This integral may be viewed as the open-string index between two 8-branes, i.e.~the structure sheaf $\mathcal{O}_{X_4}$, see \cite{Gerhardus:2016iot}. It computes the arithmetic genus of $X_4$ given by $\sum_p (-1)^p h^{0,p}$, which equals two when $X_4$ has SU(4) holonomy and not a subgroup thereof, i.e.~$h^{0,0}=h^{0,4}=1$ and $h^{0,1}=h^{0,2}=h^{0,3}=0$.}
\begin{equation}\label{eq:openstringD8}
    \frac{1}{720}\int_{X_4}\left( 3 c_2(X_4)^2-c_4(X_4)\right) = 2\, .
\end{equation}
We will use these two relations to simplify any expressions we encounter whenever convenient.

\paragraph{LCS periods and monodromy.} Let us now use the above results to express the large complex structure periods and monodromy in terms of the topological data. By acting with the transition matrix \eqref{eq:CY4T} on the Frobenius periods we find
\begin{equation}\label{eq:piLCS}
    \Pi_{\rm LCS} = \cT_{\rm LCS} \begin{pmatrix}
        1 \\
        t \\
        t^2 \\
        t^3 \\
        t^4
    \end{pmatrix} =  \left(
\begin{array}{c}
 1 \\
 -t \\
 -\frac{1}{2} t^2-\frac{1}{2}t +\frac{c_2}{24 \kappa }\\
 \frac{\kappa  }{6}t^3+\frac{\kappa  }{4}t^2+\frac{\kappa 
   }{8}t+\frac{i c_3 \zeta (3)}{8 \pi ^3}-\frac{c_2}{48} \\
 \frac{\kappa  }{24}t^4 +\frac{c_2 }{48}t^2+\frac{i c_3 t \zeta (3)}{8 \pi ^3}-\frac{c_4}{3456}-\frac{5}{12} \\
\end{array}
\right)\, .
\end{equation}
Sending $t \to t+1$ induces the monodromy transformation $\Pi(t+1) = M_0 \cdot \Pi(t)$, which reads
\begin{equation}
    M_0 = \left(
\begin{array}{ccccc}
 1 & 0 & 0 & 0 & 0 \\
 -1 & 1 & 0 & 0 & 0 \\
 -1 & 1 & 1 & 0 & 0 \\
 \frac{1}{24} \left(c_2+13 \kappa \right) & -\frac{\kappa }{2} & -\kappa  & 1 & 0 \\
 \frac{1}{24} \left(c_2+\kappa \right) & -\frac{1}{24} \left(c_2+\kappa \right) & 0 & 1 & 1 \\
\end{array}
\right)\, .
\end{equation}
Alternatively it can be obtained by conjugating \eqref{eq:Mfrob}  as $M_0 = \cT_{\rm LCS} \hat{M}_0 (T_{\rm LCS})^{-1} $. Note that integrality of the monodromy matrix $M_0 \in SO(3,2; \bbZ)$ requires the quantization condition
\begin{equation}\label{eq:quantization}
    a=\frac{\kappa+c_2}{24} \in \bbZ\, ,
\end{equation}
where we identified the monodromy parameter $a$ from before in \eqref{eq:M0}. In appendix \ref{app:Ktheory} we discuss how the LCS periods given here match with results coming from K-theory \cite{Gerhardus:2016iot,Cota:2017aal, Marchesano:2021gyv}.

\subsection{Fundamental periods, CICYs, and fibrations}\label{ssec:CICYs}
We next take a closer look at the holomorphic solutions to the Picard-Fuchs equation near $z=0$. From the expansion of this fundamental period we are able to read off the mirror Calabi--Yau fourfolds as complete intersections in weighted projective spaces. We also give a correspondence with the 14 hypergeometric Calabi--Yau threefolds of \cite{Doran:2005gu} and some related fibration structure.

\paragraph{Fundamental periods.} The fundamental periods in the large complex structure regime can be given as hypergeometric ${}_5F_4$ functions in the complex structure modulus. We may write the expansion in the large complex structure regime $\mu z \ll 1$ as
\begin{equation}\label{eq:5F4}
    \Pi^0(z) = {}_5 F_4(a_1,\ldots, a_5; 1^4; \mu z) = \sum_{n=0}^\infty c_{a_1,\ldots, a_5}(n) z^n\, ,
\end{equation}
where $\mu$ is fixed by \eqref{eq:mu}. The coefficients $c(a_1,\ldots,a_5,n)$ can be expressed in terms of factorials as
\begin{equation}\label{eq:seriescoef}
\begin{aligned}
    c_{\tfrac{1}{5},\tfrac{2}{5},\tfrac{1}{2},\tfrac{3}{5},\tfrac{4}{5}}(n) &= \frac{(2n)!(5n)!}{(n!)^7 }\, , \!  \!   &c_{\tfrac{1}{10},\tfrac{3}{10},\tfrac{1}{2},\tfrac{7}{10},\tfrac{9}{10}}(n) &= \frac{(10n)!}{(n!)^5 (5n)! }\, , \\
    c_{\tfrac{1}{2},\tfrac{1}{2},\tfrac{1}{2},\tfrac{1}{2},\tfrac{1}{2}}(n) &= \frac{(2n)!^5}{(n!)^{10} }\, , \! \!    &c_{\tfrac{1}{3},\tfrac{1}{3},\tfrac{1}{2},\tfrac{2}{3},\tfrac{2}{3}}(n) &= \frac{(3n)!^2(2n)!}{(n!)^8 }\, , \\
    c_{\tfrac{1}{3},\tfrac{1}{2},\tfrac{1}{2},\tfrac{1}{2},\tfrac{1}{3}}(n) &= \frac{(2n)!^3(3n)!}{(n!)^{9} }\, , \! \!    &c_{\tfrac{1}{4},\tfrac{1}{2},\tfrac{1}{2},\tfrac{1}{2},\tfrac{3}{4}}(n) &= \frac{(2n)!^2(4n)!}{(n!)^8 }\, , \\
    c_{\tfrac{1}{8},\tfrac{3}{8},\tfrac{1}{2},\tfrac{5}{8},\tfrac{7}{8}}(n) &= \frac{(2n)!(8n)!}{(n!)^{6}(4n)! }\, , \! \!    &c_{\tfrac{1}{6},\tfrac{1}{3},\tfrac{1}{2},\tfrac{2}{3},\tfrac{5}{6}}(n) &= \frac{(6n)!}{(n!)^6 }\, , \\
    c_{\tfrac{1}{12},\tfrac{5}{12},\tfrac{1}{2},\tfrac{7}{12},\tfrac{11}{12}}(n) &= \frac{(2n)!^2(12n)!}{(n!)^{6}(4n)!(6n)! }\, , \!  \!   &c_{\tfrac{1}{4},\tfrac{1}{4},\tfrac{1}{2},\tfrac{3}{4},\tfrac{3}{4}}(n) &= \frac{(4n)!^2}{(n!)^6(2n)! }\, , \\
    c_{\tfrac{1}{4},\tfrac{1}{3},\tfrac{1}{2},\tfrac{2}{3},\tfrac{3}{4}}(n) &= \frac{(3n)!(4n)!}{(n!)^{7}}\, , \!  \!   &c_{\tfrac{1}{6},\tfrac{1}{4},\tfrac{1}{2},\tfrac{3}{4},\tfrac{5}{6}}(n) &= \frac{(4n)!(6n)!}{(n!)^5 (2n)!(3n)! }\, , \\
    c_{\tfrac{1}{6},\tfrac{1}{6},\tfrac{1}{2},\tfrac{5}{6},\tfrac{5}{6}}(n) &= \frac{(6n)!^2}{(n!)^{4}(2n)!(3n)!^2}\, , \! \!    &c_{\tfrac{1}{6},\tfrac{1}{2},\tfrac{1}{2},\tfrac{1}{2},\tfrac{5}{6}}(n) &= \frac{(2n)!^2(6n)!}{(n!)^7 (3n)!}\, . \\
\end{aligned}
\end{equation}
We note that the usual methods of \cite{Hosono:1993qy} for obtaining the other periods also apply here. In this approach one analytically continues the factorials to Gamma-functions, replaces $n\to n+\rho$ in \eqref{eq:5F4} and takes derivatives with respect to this auxiliary variable $\rho$ before setting it to zero.

\paragraph{CICYs.} We now read off what are the mirror Calabi--Yau fourfolds from the series coefficients given in \eqref{eq:seriescoef}. Typically this method of \cite{Hosono:1993qy} is used the other way around, where one picks the mirror Calabi--Yau manifold and then determines the series expansion from the configuration matrix, but it works in reverse too. Since we are interested in the one-modulus case, the complete intersection is embedded in a single ambient space factor $\bbP^{r}[w_1 \ldots, w_{r+1}]$, where $w_1,\ldots,w_{r+1}$ are the weights of the coordinates. The CICY is then represented by a configuration matrix as
\begin{equation}
    \left(\begin{array}{c | c c c }
        \bbP^{r}[w_1 \ldots, w_{r_1+1}] & d_1 & \ldots & d_l \\
    \end{array}\right)
\end{equation}
where the $d_1,\ldots, d_l$ denote the degrees of the hypersurface polynomials. The Calabi--Yau condition, i.e.~requiring a vanishing first Chern class, corresponds to the degrees adding up to $d_1+\ldots + d_l = r+1$. The series coefficient of the fundamental period is given by
\begin{equation}\label{eq:CICYcn}
    c(n) = \frac{\prod_{j=1}^l (d_j n )!}{\prod_{j=1}^{r+1} (w_j n)!}\, .
\end{equation}
This means that we can simply read of the degrees $d_1,\ldots, d_l$ from the numerators in \eqref{eq:seriescoef}, while the weights follow from the denominators in there. These results have been included in table \ref{table:data} from the previous subsection. In turn, with all this configuration data at hand, it is possible to determine the topological data --- the intersection number and integrated Chern classes --- of this CICY by following the computational procedures of \cite{Hosono:1993qy}. While we do not explicitly write these general expressions here, we mention that their application to our fourteen fourfolds reproduces the intersection number \eqref{eq:ksine} and Chern classes \eqref{eq:chernclasses} given earlier.

\paragraph{Singular Calabi--Yau fourfold.} One of the fourteen ambient spaces listed in table \ref{table:data}, $\bbP^7[1^6,4,6]$, is singular, because two of the weights share a common divisor. The resulting Calabi--Yau fourfold $X_{2,2,12}[1^6,4,6]$ is therefore singular as well. It is instructive to look at the fourteen Calabi--Yau threefolds: among these $X_{2,12}(1^4,4,6)$ is singular for the same reason. In \cite{Klemm:2004km} this issue was circumvented by considering a blow-up of the ambient space $\bbP^5[1^4,4,6]$, which gave rise to a smooth threefold  with $h^{1,1}=3$ instead of $h^{1,1}=1$. In this work we are interested in the mirror dual side, as we want to study variations of Hodge structure over the complex structure moduli space. For the Calabi--Yau threefold this was investigated in \cite{clingher201614th}, identifying a K3 fibration with $h^{2,1}=3$ that has a matching one-parameter subvariation of Hodge structure. It would be interesting to see how all these results for threefolds generalize to the fourfold setting, but we leave this for future work. 

\paragraph{Relation to Calabi--Yau threefolds.} The fourteen Calabi--Yau fourfolds we found in this work are related in a natural manner to the fourteen hypergeometric Calabi--Yau threefolds of \cite{Doran:2005gu}. This can be observed in several ways. First of all, we note that the exponents $a_i$ of the singularities at infinity correspond one-to-one
\begin{equation}
    \text{CY3 with }(a_1,a_2,a_3,a_4) \quad \iff \quad \text{CY4 with }(a_1,a_2,\tfrac{1}{2}, a_3, a_4)\, .
\end{equation}
Also for the fundamental periods there is a straightforward correspondence, known as a Hadamard product. In fact, the hypergeometric differential operators we studied in this work were already found in this way from the threefolds in \cite{Doran:2015xjb}. A Hadamard product between two series expansions $f=\sum f_n z^n$ and $g=\sum g_n z^n$ is given by
\begin{equation}
    (f\star g)(x) = \sum_{n=0}^\infty f_n g_n z^n\, .
\end{equation}
Similar to the fundamental periods of fourfolds given in \eqref{eq:5F4}, we may give the fundamental periods of the threefolds by hypergeometric ${}_4F_3$-functions. We refer to appendix \ref{app:threefolds} for an overview on the threefold periods. We can then obtain the fourfold periods by taking the Hadamard product of these threefold periods with
\begin{equation}
    {}_1 F_0(\tfrac{1}{2}|4z) = \frac{1}{\sqrt{1-4z}} =  \sum_{n=0}^\infty \frac{(2n)!}{(n!)^2} z^n\, .
\end{equation}
To be precise, this Hadamard product is identified with the fourfold period as
\begin{equation}
    {}_5 F_4(a_1,a_2, \tfrac{1}{2},a_3, a_4; 1^4| \mu_4 z) = {}_1 F_0(\tfrac{1}{2}|4z) \star {}_4 F_3(a_1,\ldots, a_4; 1^3| \mu_3 z)
\end{equation}
where $\mu_4 = 4 \mu_3$. We can also write out this identity at the level of the series coefficients as
\begin{equation}\label{eq:series34}
    c_{a_1,a_2,\frac{1}{2},a_3,a_4}(n) = \frac{(2n)!}{(n!)^2} c_{a_1,a_2,a_3,a_4}(n)\, ,
\end{equation}
where $c_{a_1,a_2,a_3,a_4}(n)$ denote the series coefficients of ${}_4 F_3(a_1,\ldots, a_4; 1^3| \mu_3 z)$. From the perspective of the CICY data this can be understood as adding two coordinates with weight $w_{r+2}=w_{r+2}=1$ to the ambient space, while intersecting with an additional hypersurface polynomial of degree $d_{l+1}=2$. Note, however, that the factor $(2n)!$ cancels against a factor in the threefold coefficient $c_{a_1,a_2,a_3,a_4}(n)$ when this CICY has a coordinate with weight $w=2$ in its ambient space. In this case, instead of adding another degree two polynomial, this coordinate is simply removed from the weighted projective space.

\paragraph{Fibration structure.} While the above discussion about CICYs concerns the mirrors, the Calabi--Yau fourfolds themselves can also be related to lower-dimensional Calabi--Yau manifolds. In fact, this fibration structure has already been made precise in \cite{Doran:2015xjb}, where an iterative construction was defined that produces elliptically fibered Calabi--Yau manifolds from elliptic Calabi--Yau manifolds of one dimension lower. This method produced fourteen Calabi--Yau fourfolds as elliptic fibrations over $\bbP^1\times \bbP^1\times \bbP^1$ together with the Picard-Fuchs equations for the periods. These are precisely the fourteen Calabi--Yau fourfolds considered in this paper. Compared to \cite{Doran:2015xjb} one of the main new contributions of our classification is arguing that these are all Calabi--Yau fourfolds with $\cM_{\rm cs}=\bbP^1\backslash\{0,1,\infty\}$, as well as setting up the integral basis for the periods.

\subsection{Main example: \texorpdfstring{$X_6$}{X6}}\label{ssec:X6}
To illustrate our analysis above, let us consider the sextic $X_6$ as an illustrative example. We take here the perspective as if nothing about the geometry is known to us, and reconstruct this information ourselves from analysis of the monodromies and periods.

\paragraph{Monodromies.} Let us begin from the monodromies. For the sextic we want the monodromy matrix given in \eqref{eq:Minfty} to be of finite order 6, i.e.~it satisfies $(M_\infty)^6=1$. We can solve this equation for $\kappa,a$ over the integers: the unique solution is $\kappa=6$ and $a=4$. Let us record the resulting monodromies around $0$ and $\infty$ for completeness
\begin{equation}\label{eq:sexticMs}
    M_0 = \left(
\begin{array}{ccccc}
 1 & 0 & 0 & 0 & 0 \\
 -1 & 1 & 0 & 0 & 0 \\
 -1 & 1 & 1 & 0 & 0 \\
 7 & -3 & -6 & 1 & 0 \\
 4 & -4 & 0 & 1 & 1 \\
\end{array}
\right)\, , \ \qquad \ M_\infty = \left(
\begin{array}{ccccc}
 -4 & -7 & 6 & 1 & -1 \\
 1 & 1 & 0 & 0 & 0 \\
 0 & -1 & 1 & 0 & 0 \\
 -4 & -3 & 6 & 1 & 0 \\
 -1 & 0 & 0 & 0 & 0 \\
\end{array}
\right)\, ,
\end{equation}
with the monodromy around the conifold point given by \eqref{eq:MC} as in all examples. 

\paragraph{Picard-Fuchs equation and Frobenius solution.} From the eigenspectrum of the monodromy we can determine the differential equation for the periods. With the local exponents given in terms of the eigenvalues of $M_\infty$ by \eqref{eq:expsdef}, we find from \eqref{eq:PFfourfold} for the sextic
\begin{equation}
    L_{\rm sextic} = \theta^5 -6^{-6} z(\theta+\tfrac{1}{6})(\theta+\tfrac{1}{3})(\theta+\tfrac{1}{2})(\theta+\tfrac{2}{3})(\theta+\tfrac{5}{6})\, ,
\end{equation}
where we recall that $\theta = z \frac{d}{dz}$. We now take the Frobenius ans\"atze \eqref{eq:frobLCS} and \eqref{eq:frobC} and solve order-by-order for the coefficients. For completeness we record here the first few orders, while in the accompanying notebook we went up to order 100.  For $z=0$ we have the power series
\begin{equation}
\begin{aligned}
    f_0(z) &= 1 +720 z + 7484400 z^2 + 137225088000 z^3 + \mathcal{O}(z^4) \, , \\
    f_1(z) &= 6264 z+ 71994420 z^2 +1368223113600 z^3 +  \mathcal{O}(z^4) \, , \\
    f_2(z) &= 20160 z+ 327001536 z^2 +6903566901440 z^3 +  \mathcal{O}(z^4)  \, , \\
    f_3(z) &= -60480 z-111585600 z^2 +3717865540864 z^3 +  \mathcal{O}(z^4) \, , \\
    f_4(z) &= -2734663680 z^2 -57797926824960 z^3 +  \mathcal{O}(z^4)  \, . \\
\end{aligned}
\end{equation}
while for the conifold point at $z=6^{-6}$ we find as Frobenius periods
\begin{equation}
\begin{aligned}
    \varpi_0(z) &= 1 -1218719480020992 z^4 +\tfrac{849749720005676630016}{7} z^5+ \mathcal{O}(z^6)   \, , \\
    \varpi_1(z) &= z- \tfrac{51746469691392}{5} z^4+ \tfrac{4752125785790742528}{5} z^5 +\mathcal{O}(z^6) \, , \\
    \varpi_{\frac{3}{2}}(z) &= z^\frac{3}{2}-44064z^{\frac{5}{2}} +1850936832 z^{\frac{7}{2}}+ \mathcal{O}(z^{\frac{9}{2}}) \, , \\
    \varpi_2(z) &=  z^2-1859334912z^4+\tfrac{751029088002048}{5}z^5+ \mathcal{O}(z^6) \, , \\
    \varpi_3(z) &= z^3 -81000 z^4 +4874245632 z^5+ \mathcal{O}(z^6)\, . 
\end{aligned}
\end{equation}

\paragraph{Monodromies in LCS Frobenius basis.} Having set up the Frobenius periods around the LCS and conifold point, let us next bring both monodromies to the same basis. By expanding both sets of periods around the midpoint $z_{\rm mid}= 2^{-7} 3^{-6}$ we find as transition matrix
\begin{equation}
     \cT_{\rm C}^{\ \rm LCS} = \scalebox{0.9}{$\left(
\begin{array}{ccccc}
 1.02252 & 2407.32 & 589523 i & -1.25693\times 10^8 & 5.32632\times 10^{12} \\
 1.71697 i &  -7235.28 i & 0 & 1.71464\times 10^8 i & -5.35457\times 10^{12} i \\
 -2.90188 & 28378.9 & 736904 i & -8.03988\times 10^8 & 2.76423\times 10^{13}
   \\
 4.94221 i & 70815.7 i & -1.19987\times 10^6 & -2.0664\times 10^9 i
   & 7.18871\times 10^{13} i \\
 8.48867 & -152282 &  -405297 i & 4.54756\times 10^9 & -1.60062\times 10^{14}
   \\
\end{array}
\right)$}\, ,
\end{equation}
where we recorded the first six digits for brevity, and in the notebook we continued up to 50 digits. We can use this transition matrix to bring the conifold monodromy in \eqref{eq:Mfrob} to the LCS Frobenius basis as
\begin{equation}
    \cT_{\rm C}^{\ \rm LCS} \hat{M}_{\rm C} (\cT_{\rm C}^{\ \rm LCS})^{-1} = \left(
\begin{array}{ccccc}
 1.17188 & 2.03533 i & -1.875 & 0 & -0.25 \\
 0 & 1 & 0 & 0 & 0 \\
 0.214844 &  2.54416 i & -1.34375 & 0 & -0.3125 \\
 0.349822 i & -4.14257 & -3.81624 i & 1 & -0.508832 i \\
 -0.118164 & -1.39929 i & 1.28906 & 0 & 1.17188 \\
\end{array}
\right)\, .
\end{equation}
The question is now to find a transition matrix that rotates this conjugated monodromy matrix and $\hat{M}_0$ in \eqref{eq:Mfrob} simultaneously into \eqref{eq:sexticMs}. This is the unique basis transformation described in \eqref{eq:uniqueB}, which is found to be
\begin{equation}
    \cT_{\rm sextic} = \left(
\begin{array}{ccccc}
 1 & 0 & 0 & 0 & 0 \\
 0 & -1 & 0 & 0 & 0 \\
 0.625 & -0.5 & -0.5 & 0 & 0 \\
 -1.875-2.03533 i & 0.75 & 1.5 & 1 & 0 \\
 -1.17188 &  -2.03533 i & 1.875 & 0 & 0.25 \\
\end{array}
\right)
\end{equation}
And indeed, upon comparing with \eqref{eq:CY4T} we notice that the two agree numerically for the topological data
\begin{equation}
    \kappa=6\, ,\qquad c_2 = 90\, , \qquad c_3 = -420\, .
\end{equation}
These are precisely the topological numbers known for the sextic. We have iterated this scan over the complete set of fourteen Calabi--Yau fourfolds, resulting in the topological data given in table \ref{table:data}. In all cases this topological data matches with the sine and polygamma identities given in \eqref{eq:ksine} and \eqref{eq:c2c3}. We have also cross-checked with the CICYs we identified, and this topological data matches with what is obtained from the standard expressions of \cite{Hosono:1993qy}.

\section{Periods at infinity}\label{sec:infmonodromies}
In this section we study the four types of singularities arising at infinity: LCS points, CY3-points, C-points and F-points. For each type we consider a working example where we write down the data that captures the physics close to these singularities. Our study covers the monodromy matrix, leading periods and K\"ahler potential, and the limiting mixed Hodge structure. These models can be thought of as complementary to the threefold limits studied in \cite{Bastian:2021hpc, Bastian:2023shf}, and the two-moduli limits of Calabi--Yau fourfolds studied in \cite{Grimm:2019ixq}.

\subsection{LCS point}\label{ssec:LCS}
We first consider the case where there is another large complex structure point at infinity. These singularities arise when all local exponents of the periods are equal, i.e.~$a_1=\ldots=a_5$. This only happens for case \#3 in table \ref{table:data} with exponents $(\tfrac{1}{2},\tfrac{1}{2},\tfrac{1}{2},\tfrac{1}{2},\tfrac{1}{2}) $, which we therefore consider as working example here. Geometrically this case corresponds to the mirror of the Calabi--Yau fourfold realized as the intersection of five quadrics in $\bbP^9$.

\paragraph{Basis transformation.} Before we get into the details of this example, let us fix a basis where the boundary data takes a simpler form. We consider the unimodular basis transformation
\begin{equation}
    B = \left(
\begin{array}{ccccc}
 0 & 0 & 1 & 0 & -4 \\
 0 & 0 & 0 & 1 & 2 \\
 0 & 0 & 0 & 1 & 1 \\
 1 & 2 & 4 & -16 & -16 \\
 0 & 1 & 1 & -4 & -4 \\
\end{array}
\right)\, , 
\end{equation}
which renders the monodromy matrix lower-triangular. To be precise, $M_\infty $ transforms as $M_\infty \to B^{-1} M_\infty B$ and the pairing as $\eta \to B^T \eta B$, which yields
\begin{equation}\label{eq:LCSdata}
    M_\infty = \left(
\begin{array}{ccccc}
 -1 & 0 & 0 & 0 & 0 \\
 -1 & -1 & 0 & 0 & 0 \\
 1 & 1 & -1 & 0 & 0 \\
 0 & 0 & -1 & -1 & 0 \\
 0 & 0 & 1 & 1 & -1 \\
\end{array}
\right)\, , \qquad \eta = \left(
\begin{array}{ccccc}
 0 & 0 & 0 & 1 & 2 \\
 0 & 0 & 1 & 2 & 0 \\
 0 & 1 & 2 & 0 & 0 \\
 1 & 2 & 0 & 0 & 0 \\
 2 & 0 & 0 & 0 & 0 \\
\end{array}
\right)\, .
\end{equation}
Notice that the semi-simple part of $M_\infty$ is just minus the identity matrix. 

\paragraph{Mirror of second LCS point.} Recall that mirror symmetry states that every LCS regime should be identified with the large volume regime of a mirror dual Calabi--Yau manifold. So when we have multiple LCS points, each LCS point has a mirror dual Calabi--Yau manifold. In \cite{Schimannek:2021pau, Katz:2022lyl} this correspondence was worked out for a Calabi--Yau threefold with two LCS points, i.e.~operator AESZ 3 of \cite{almkvist2005tables}: the first mirror is the intersection of four quadrics in $\bbP^7$, while the other is (the non-commutative resolution of) the octic in $\bbP^4[1^4,4]$. Although we do not attempt to identify the mirror fourfold of our second LCS point explicitly, we still want to draw some lessons from the period data available to us. In \cite{Schimannek:2021pau, Katz:2022lyl} it was explained that we should consider the square of $M_\infty$ to match with the mirror dual, as this removes the semisimple part:
\begin{equation}\label{eq:MLCS}
    M_{\rm LCS}  = (M_\infty)^2 =\left(
\begin{array}{ccccc}
 1 & 0 & 0 & 0 & 0 \\
 2 & 1 & 0 & 0 & 0 \\
 -3 & -2 & 1 & 0 & 0 \\
 -1 & -1 & 2 & 1 & 0 \\
 1 & 1 & -3 & -2 & 1 \\
\end{array}
\right)\, .
\end{equation}
Then we notice some key differences compared to the standard LCS  monodromy \eqref{eq:M0}. What stands out immediately are the 2's instead of 1's in entries $(2,1)$ and $(5,4)$, which cannot be removed by an integral basis transformation. Similarly, the 1's in the pairing in \eqref{eq:LCSdata} cannot be removed while preserving the lower-triangular form of $M_\infty$. Both of these observations tell us that the ansatz \eqref{eq:M0} we made for the large complex structure monodromy should be revised if we want to go beyond the models discussed in this work.

\paragraph{Asymptotic periods.} Let us now expand the periods at this second large complex structure point. As coordinate we take a covering coordinate $t$ defined by
\begin{equation}
    z = 2^{-20}e^{-4\pi i t}
\, ,
\end{equation}
such that $z=\infty$ corresponds to $t=i\infty$. We included the overall factor to remove all $\log 2$ terms from the periods. Circling the singularity by $t \to t+1$ corresponds to applying the monodromy $M_\infty$ twice, so we work instead with $M_{\rm LCS}$ in \eqref{eq:MLCS}. We normalize our periods in accordance with \cite{Katz:2022lyl}, where the precise mirror correspondence for a Calabi--Yau threefold with a second large complex structure point was worked out. Including only polynomial terms in $t$ and suppressing all exponential corrections, we find
\begin{equation}\label{eq:piLCSinf}
    \Pi(t) = \left(
\begin{array}{c}
 \frac{1}{2} \\
 t \\
 -t^2-\frac{t}{2}-\frac{1}{48} \\
 -\frac{2 t^3}{3}+\frac{t}{8}+\frac{5 i \zeta (3)}{8 \pi ^3} \\
 \frac{t^4}{3}+\frac{t^3}{3}-\frac{t^2}{24}-\frac{5 i t \zeta (3)}{4 \pi ^3}-\frac{t}{16}-\frac{5 i \zeta (3)}{16 \pi ^3}-\frac{1}{2304} \\
\end{array}
\right) + \mathcal{O}(e^{2\pi i t})\, .
\end{equation}
Note in particular the different normalization for the first period, which conspires with the monodromy matrix in \eqref{eq:MLCS} such that we still shift $t \to t+1$ under monodromies. The corresponding K\"ahler potential reads
\begin{equation}
    e^{-K} = \frac{16 s^4}{3}+\frac{5 s \zeta (3)}{\pi ^3} + \mathcal{O}(e^{-2\pi s})\, .
\end{equation}
where we expanded $t = a+i s$, with the axion $a$ dropping out at polynomial order.

\paragraph{Limiting mixed Hodge structure.} Let us finally consider the limiting mixed Hodge structure associated to this singularity. We refer to appendix \ref{app:lmhs} for a brief review to this topic, and just list the main results we found here. The dimensions of the splitting are given by
\begin{equation}
    \dim I^{p,q} = \begin{tikzpicture}[baseline={([yshift=-.5ex]current bounding box.center)},scale=0.5,cm={cos(45),sin(45),-sin(45),cos(45),(15,0)}]
\draw (0,0) node {$1$};
\draw (1,0) node {$0$};
\draw (2,0) node {$0$};
\draw (3,0) node {$0$};
\draw (4,0) node {$0$};

\draw (0,1) node {$0$};
\draw (1,1) node {$1$};
\draw (2,1) node {$0$};
\draw (3,1) node {$0$};
\draw (4,1) node {$0$};

\draw (0,2) node {$0$};
\draw (1,2) node {$0$};
\draw (2,2) node {$1$};
\draw (3,2) node {$0$};
\draw (4,2) node {$0$};

\draw (0,3) node {$0$};
\draw (1,3) node {$0$};
\draw (2,3) node {$0$};
\draw (3,3) node {$1$};
\draw (4,3) node {$0$};

\draw (0,4) node {$0$};
\draw (1,4) node {$0$};
\draw (2,4) node {$0$};
\draw (3,4) node {$0$};
\draw (4,4) node {$1$};
\end{tikzpicture}\,  .
\end{equation}
The top component of the limiting mixed Hodge structure is spanned by the vector
\begin{equation}
\begin{aligned}
    I^{4,4}: \left(\tfrac{1}{2},0,-\tfrac{1}{48},\tfrac{5 i \zeta (3)}{8 \pi ^3},-\tfrac{1}{2304}-\tfrac{5 i \zeta (3)}{16 \pi ^3}\right)\, .
\end{aligned}
\end{equation}
Note that this is precisely the constant part of the period vector written in \eqref{eq:piLCSinf}. Its descendants $I^{3,3}, I^{2,2}, I^{1,1}$, and $ I^{0,0}$ may be obtained by acting with the log-monodromy operator
\begin{equation}
    N_\infty = \log[M_{\rm LCS}] = \scalebox{0.9}{$\left(
\begin{array}{ccccc}
 0 & 0 & 0 & 0 & 0 \\
 2 & 0 & 0 & 0 & 0 \\
 -1 & -2 & 0 & 0 & 0 \\
 \frac{1}{3} & 1 & 2 & 0 & 0 \\
 -\frac{1}{6} & -\frac{1}{3} & -1 & -2 & 0 \\
\end{array}
\right)$}\, ,
\end{equation}
so we refrain from writing their basis vectors here.
\subsection{CY3 points}
Here we consider CY3-points, where the first and last pair of local exponents are equal: $a_1 =a_2$ and $a_4=a_5$, but $a_2 < a_3 <a_4$. We call it a CY3-point because the asymptotic periods contain the data of a rigid Calabi--Yau threefold; see also the diamond of the limiting mixed Hodge structure in \eqref{eq:CY3diamond}. This type of singularity arises in three cases in table \ref{table:data}: \#4, \#10 and \#13. As working example we take \#13 with local exponents $(\tfrac{1}{6},\tfrac{1}{6},\tfrac{1}{2},\tfrac{5}{6},\tfrac{5}{6})$, corresponding to the mirror of the intersection of two sextics in $\bbP^6[1^4,2,3^2]$. 

\paragraph{Analogy with K-points.} We first draw a comparison to Calabi--Yau threefolds, where the analogue of a CY3-point is a K-point. The geometries arising at these K-points are well-understood in the mathematics literature; for instance, the threefold cousins of our three fourfolds \#4, \#11 and \#13 have been studied in \cite{Doran:2016uea}. We give some intuition for what happens at these singular points here, and refer to \cite{Doran:2016uea} for the details. On the complex structure side a K-point corresponds to Tyurin degeneration \cite{tyurin}, a particular kind of semi-stable degeneration limit. Namely, the Calabi--Yau threefold degenerates into the union of two components that intersect along a K3 surface with maximal Picard rank, i.e.~without any complex structure moduli. On the mirror dual side these limits correspond to K3 fibrations whose base volume is made large. These mirrors have $h^{1,1}=1$, so while this appears to forbid any fibration structure, this is reconciled with the fact that the K3 surface is non-commutative \cite{Katzarkov}. Coming to our Calabi--Yau fourfolds, we expect that a similar picture carries over on both sides, with the K3 surface replaced by a rigid/non-commutative Calabi--Yau threefold. For the remainder of this section we put these intricaties aside and leave them to future investigations.

\paragraph{Physical interpretation.} In the F-theory literature we know semi-stable degeneration limits as Sen limits \cite{Sen:1996vd, Clingher:2012rg}. Physically it describes the weak-coupling limit of F-theory to Type IIB orientifolds. Since our Calabi--Yau fourfolds have a single complex structure modulus, this means that this modulus parametrizes the string coupling, and our Type IIB background is given by some rigid Calabi--Yau orientifold. These CY3-points thus provide us with a setting where we have explicit control over non-perturbative corrections coming from D($-1$)-instantons, as we can compute the fourfold periods numerically to a very high degree, and even give a closed form in terms of hypergeometric functions. Furthermore, in addition to the usual periods expected in this rigid Type IIB orientifold, there is an additional linear combination of periods that is exponentially small in the string coupling. From \cite{Clingher:2012rg, Alim:2009bx, Grimm:2009ef,Jockers:2009ti} we expect this to come from D7-branes in the Type IIB orientifold, where worldvolume fluxes can induce a superpotential. In this paper we work directly with the fourfold periods, but it would be interesting to work out the D7-brane picture more precisely.

\paragraph{Monodromy matrix.} Having given some geometrical and physical context to CY3-points, we now consider the example at hand. In order to simplify the expressions, we consider the unimodular basis transformation
\begin{equation}
B = \scalebox{0.9}{$\left(
\begin{array}{ccccc}
 -1 & 1 & 2 & 0 & 1 \\
 0 & 0 & 0 & -1 & -1 \\
 0 & 0 & 1 & -1 & 0 \\
 -1 & 0 & 0 & 1 & 0 \\
 0 & 0 & 0 & 1 & 0 \\
\end{array}
\right)$} \, .
\end{equation}
This is an isometry of the pairing, so $\eta$ is still given by \eqref{eq:eta} with $\kappa=2$. The monodromy matrix transforms as $M_\infty \to B^{-1} M_\infty B$, which decomposes into semisimple and unipotent factors
\begin{equation}\label{eq:CY3monodromy}
    M_{ss} = \scalebox{0.9}{$\left(
\begin{array}{ccccc}
 0 & 1 & 0 & 0 & 0 \\
 -1 & 1 & 0 & 0 & 0 \\
 1 & -1 & -1 & 0 & 0 \\
 1 & -\frac{4}{3} & -2 & 0 & -1 \\
 -\frac{1}{3} & \frac{1}{3} & 0 & 1 & 1 \\
\end{array}
\right)$}\, , \qquad M_u = \scalebox{0.9}{$\left(
\begin{array}{ccccc}
 1 & 0 & 0 & 0 & 0 \\
 0 & 1 & 0 & 0 & 0 \\
 0 & 0 & 1 & 0 & 0 \\
 2 & 0 & 0 & 1 & 0 \\
 0 & -2 & 0 & 0 & 1 \\
\end{array}
\right)$}\, .
\end{equation}
The semisimple factor is of order six, and has eigenvalues $e^{\pi i/3}$, $-1$, and $e^{-\pi i/3}$; the eigenvectors are given by the limiting mixed Hodge structure in \eqref{eq:CY3lmhs}. The unipotent factor has been reduced to a single $2\times 2$ subblock. We stress that, while this is also possible for the other two CY3-points, the pairing \eqref{eq:eta} will also get off-diagonal entries from these basis transformations.

\paragraph{Asymptotic period vector.} We now expand the period vector in the vicinity of the CY3-point. We do so in the coordinate
\begin{equation}
    z \to -\frac{1}{2^{20}3^9 z}\, ,
\end{equation}
such that the CY3-point lies at $z=0$. The additional factors are included to remove all $\log[2]$ and $\log[3]$ terms from the periods. Note that we expect the string coupling to be related to this complex structure coordinate as $\tau \sim \log[z]/2\pi i$. In terms of $z$ the period vector reads
\begin{equation}\label{eq:CY3pi}
\begin{aligned}
    \mathbf{\Pi} &= A z^{1/6} \begin{pmatrix}
        1\\
        \tfrac{1}{2}+\tfrac{i\sqrt{3}}{2}\\
        -\tfrac{i}{\sqrt{3}}\\
        \frac{\log[z]}{6\pi i}-\tfrac{2i}{3\sqrt{3}}\\
        \tfrac{i}{3\sqrt{3}}-(\frac{1}{2}+\frac{i\sqrt{3}}{2})\frac{\log[z]}{6\pi i}
    \end{pmatrix}- \frac{3 i }{2^{1/3}\pi^2 }  z^{1/2}\begin{pmatrix}
        0 \\
        0 \\
        3 \\
        4 \\
        -2 \\
    \end{pmatrix}\\
    & \ \ \ - \frac{27 \sqrt{3} }{2^{2/3}\pi^3 A }z^{5/6} \begin{pmatrix}
        1\\
        \tfrac{1}{2}-\tfrac{i\sqrt{3}}{2}\\
        \tfrac{i}{\sqrt{3}}\\
        \frac{\log[z]-6}{6\pi i}+\tfrac{2i}{3\sqrt{3}}\\
        (\tfrac{1}{2}-\tfrac{i\sqrt{3}}{2})\frac{\log[z]-6}{6\pi i}-\tfrac{i}{3\sqrt{3}}
    \end{pmatrix} +\mathcal{O}(z^{7/6})\, ,
\end{aligned}
\end{equation}
where apart from $\pi$ the only other non-algebraic number appearing is
\begin{equation}
    A = \frac{\Gamma \left(\frac{1}{6}\right)^4 \Gamma \left(\frac{1}{3}\right)}{32 \sqrt{3} \pi ^4}\, \simeq 0.476348\, .
\end{equation}
From these periods we compute the K\"ahler potential
\begin{equation}
e^{-K} = - \frac{A^2}{\sqrt{3}\pi}|z|^{1/3} \log |z| +\frac{3^4 2^{1/3}|z|}{\pi^4}  +\frac{729 \sqrt{3}}{ 2^{4/3} A^2 \pi^7}|z|^{5/3}(\log |z|-6) + \mathcal{O}(|z|^{4/3})\, .
\end{equation}
As the factor of $|z|^{1/3}$ may be removed by a K\"ahler transformation $\mathbf{\Pi} \to z^{-1/6}\mathbf{\Pi}$, this gives a linear behavior in the string coupling $g_s \sim \log|z|$, as expected for a weak-coupling limit. On the other hand, we also have exponential corrections $z \sim e^{-1/g_s}$ that we expect to come from D($-1$)-instantons in the Type IIB orientifold. Among these corrections we wrote down a term at order $|z|^{5/3}$, however, there are also other terms contributing below this order: the term at order $z^{1/6}$ in $\mathbf{\Pi}$ combines with a term at order $\bar z^{7/6}$ in $\mathbf{\bar \Pi}$ to give a term at order $|z|^{4/3}$. In fact, this is the first term breaking the continuous shift symmetry of the axion $\arg z$.

\paragraph{D7-brane flux superpotential.} We now turn on a four-form flux $G_4=q_{\rm D7}(0,0,3,4,-2)$, with $q_{\rm D7} \in \bbZ$ some flux quantum. As explained earlier, in the Type IIB picture we expect $q_{\rm D7}$ to be a worldvolume flux on some D7-brane. This gives a superpotential
\begin{equation}
\begin{aligned}
    W_{\rm D7} &= \frac{27 i 2^{2/3} q_{\rm D7} \sqrt{z}}{\pi ^2}\,_5F_4\left(\tfrac{1}{2}^5;\tfrac{2}{3}^2,\tfrac{4}{3}^2;-2^{10}3^3 z\right) \, ,
\end{aligned}
\end{equation}
which admits an expansion as a hypergeometric series
\begin{equation}
\begin{aligned}
    W_{\rm D7} &= \frac{ i 2^{8/3} q_{\rm D7} }{\pi ^2} \sqrt{z}\sum_{k=0}^\infty  \frac{ \Gamma \left(k+\frac{1}{2}\right)^5}{\sqrt{\pi }\Gamma(k+1) \Gamma
   \left(k+\frac{2}{3}\right)^2 \Gamma \left(k+\frac{4}{3}\right)^2} (-2^{10} 3^{3}  z)^k\\
   &= \frac{27 i 2^{2/3} q_{\rm D7} }{\pi ^2} \sqrt{z} \left( 1-\tfrac{2187 }{2}z+\tfrac{9298091736 }{1225}z^2 -\tfrac{4236443047215 }{49} z^3+ \mathcal{O}(z^4) \right)\, ,
\end{aligned}
\end{equation}
For the other two CY3-points we can write down similar D7-brane flux superpotentials. In all cases $W_{\rm D7}$ is described by the Frobenius solution proportional to $\sqrt{z}$ around $z=\infty$: the accompanying hypergeometric function for \#4 is ${}_5F_4(\tfrac{1}{2}^5;\tfrac{5}{6}^2,\tfrac{7}{6}^2;z)$ and for \#10 it is $ {}_5F_4(\tfrac{1}{2}^5;\tfrac{3}{4}^2,\tfrac{5}{4}^2;z)$. 

\paragraph{Limiting mixed Hodge structure.} Following the methods laid out in appendix \ref{app:lmhs} we next determine the limiting mixed Hodge structure. For all CY3-points the diamond looks like

\begin{equation}\label{eq:CY3diamond}
        \dim I^{p,q} = \begin{tikzpicture}[baseline={([yshift=-.5ex]current bounding box.center)},scale=0.5,cm={cos(45),sin(45),-sin(45),cos(45),(15,0)}]
\draw (0,0) node {$0$};
\draw (1,0) node {$0$};
\draw (2,0) node {$0$};
\draw (3,0) node {$1$};
\draw (4,0) node {$0$};

\draw (0,1) node {$0$};
\draw (1,1) node {$0$};
\draw (2,1) node {$0$};
\draw (3,1) node {$0$};
\draw (4,1) node {$1$};

\draw (0,2) node {$0$};
\draw (1,2) node {$0$};
\draw (2,2) node {$1$};
\draw (3,2) node {$0$};
\draw (4,2) node {$0$};

\draw (0,3) node {$1$};
\draw (1,3) node {$0$};
\draw (2,3) node {$0$};
\draw (3,3) node {$0$};
\draw (4,3) node {$0$};

\draw (0,4) node {$0$};
\draw (1,4) node {$1$};
\draw (2,4) node {$0$};
\draw (3,4) node {$0$};
\draw (4,4) node {$0$};
\end{tikzpicture}\,  .
\end{equation}
Note the two weight-three Hodge structures at rows $p+q=3$ and $p+q=5$, corresponding to the middle cohomology of a rigid Calabi--Yau threefold. We get two copies because we can expand both along the $A$ and $B$-cycle of the F-theory torus. In terms of fluxes the upstairs part corresponds to the NS-NS flux $H_3$ while downstairs to R-R flux $F_3$. For the example at hand the limiting mixed Hodge structure is spanned by
\begin{equation}\label{eq:CY3lmhs}
\begin{aligned}
    I^{4,1} &: \big(1,\tfrac{1}{2}+ \tfrac{i \sqrt{3}}{2},-\tfrac{i}{\sqrt{3}},-\tfrac{2 i}{3 \sqrt{3}},\tfrac{i}{3 \sqrt{3}}\big)\, , \quad I^{2,2}: (0, 0, 3, 4, -2)\, , \quad I^{3,0} : \big(0,0,0,1, -\tfrac{1}{2}-\tfrac{i\sqrt{3}}{2}\big)\, ,\\
\end{aligned}
\end{equation}
and $I^{1,4}$ and $I^{0,3}$ fixed as the complex conjugates. These vector spaces are eigenspaces of $M_{ss}$: $I^{4,1}$ and $I^{3,0}$ have eigenvalue $e^{2\pi i/6}$, $I^{2,2}$ has minus one, and $I^{1,4}$ and $I^{0,3}$ have $e^{-2\pi i/6}$. This splitting precisely coincides with the exponents $(\tfrac{1}{6}, \tfrac{1}{6}, \tfrac{1}{2}, \tfrac{5}{6}, \tfrac{5}{6})$ of the local period solutions, and also the terms in the period vector expansion \eqref{eq:CY3pi} are spanned by the respective eigenvectors.

\subsection{Conifold points}
In this section we consider conifold points at infinity. By conifold point we mean here that the middle three local exponents of the periods are equal: $a_2=a_3=a_4$ but $a_1<a_2<a_5$. Note that these exponents differ from the conifold points at $z=1$, which have exponents $(0,1,2,3,\frac{3}{2})$.  In table \ref{table:data} we find three conifold points at infinity: \#5, \#6 and \#14. We take \#5 as working example here, which has local exponents $(\tfrac{1}{3},\tfrac{1}{2},\tfrac{1}{2},\tfrac{1}{2},\tfrac{2}{3})$, with its mirror being $\bbP^8_{2,2,2,3}[1^9]$.

\paragraph{Monodromy matrix and pairing.} In order to simplify the monodromy matrix while keeping the pairing in a manageable form, we consider the unimodular basis transformation
\begin{equation}
    B = \scalebox{0.9}{$\left(
\begin{array}{ccccc}
 1 & 0 & 4 & 3 & 0 \\
 0 & 1 & -2 & -1 & -1 \\
 0 & 1 & -1 & 0 & -1 \\
 3 & -13 & 14 & -1 & 11 \\
 1 & -4 & 4 & 0 & 3 \\
\end{array}
\right)$} \, , \qquad \eta =  \scalebox{0.9}{$\left(
\begin{array}{ccccc}
 2 & -1 & 2 & 0 & 0 \\
 -1 & -2 & 0 & 0 & 0 \\
 2 & 0 & 0 & 0 & 0 \\
 0 & 0 & 0 & 2 & -1 \\
 0 & 0 & 0 & -1 & 2 \\
\end{array}
\right)$}\, .
\end{equation}
We recognize that the second block of $\eta$ corresponds to the pairing matrix of the $A_2$ root lattice. The monodromy matrix factorizes into a semisimple and unipotent factor
\begin{equation}\label{eq:Cmonodromy}
    M_{ss} = \scalebox{0.9}{$\left(
\begin{array}{ccccc}
 -1 & 0 & 0 & 0 & 0 \\
 0 & -1 & 0 & 0 & 0 \\
 0 & 0 & -1 & 0 & 0 \\
 0 & 0 & 0 & -1 & 1 \\
 0 & 0 & 0 & -1 & 0 \\
\end{array}
\right)$}\, , \qquad  M_u = \scalebox{0.9}{$\left(
\begin{array}{ccccc}
 1 & 0 & 0 & 0 & 0 \\
 1 & 1 & 0 & 0 & 0 \\
 1 & 1 & 1 & 0 & 0 \\
 0 & 0 & 0 & 1 & 0 \\
 0 & 0 & 0 & 0 & 1 \\
\end{array}
\right)$}\, .
\end{equation}
The semisimple monodromy is of order six, and consists of two blocks: minus the $3\times3$ identity matrix and the familiar $\bbZ_3$ generator from SL$(2,\bbZ)$.

\paragraph{Asymptotic periods.} We first perform the change of coordinate
\begin{equation}
    z \to 2^{-12} z^{-1}\, ,
\end{equation}
such that the singularity is located at $z=0$, and monodromies correspond to $z \to e^{2\pi i}z$. The rescaling by $2^{-12}$ has been included to remove any $\log[2]$ terms from the periods. In our expansion we keep the orders $z^a$ for the local exponents $a=\frac{1}{3}, \frac{1}{2}, \frac{2}{3}$ of the periods. This yields as local period vector at the CY3-point
\begin{equation}\label{eq:Cseries}
    \mathbf{\Pi} = A z^{\frac{1}{3}} \left(
\begin{array}{c}
 0 \\
 0 \\
 0 \\
 1 \\
 \tfrac{1}{2}+\tfrac{i\sqrt{3}}{2} \\
\end{array}
\right)- \frac{\sqrt{z}}{\pi i} \left(
\begin{array}{c}
1 \\
 \frac{\log[z]-6}{2\pi i} \\
 \frac{(\log[z]-6+\pi i)^2+36+5\pi^2}{2(2 \pi i)^2} \\
 0 \\
 0 \\
\end{array}
\right)-\frac{3}{\pi^4 A} z^{\frac{2}{3}} \left(
\begin{array}{c}
 0 \\
 0 \\
 0 \\
 1 \\
 \tfrac{1}{2}-\tfrac{i\sqrt{3}}{2} \\
\end{array}
\right) + \mathcal{O}(z^{\frac{4}{3}})\, ,
\end{equation}
where we defined a shorthand for the coefficient
\begin{equation}
    A = \frac{\sqrt{3}  \Gamma \left(\frac{1}{6}\right)^3 \Gamma \left(\frac{1}{3}\right)^6}{128  \pi ^{13/2}} \simeq 0.50624\, .
\end{equation}
Notice in particular the `rigid' period $\frac{1}{2}+\frac{i\sqrt{3}}{2}$ appearing in the leading term (and its conjugate at order $z^{2/3}$). It would be interesting to study how this parameter relates to the singular geometry arising in the limit, similar to how for Calabi--Yau threefolds at a conifold point the periods of the conifold show up \cite{Bonisch:2022mgw, Bastian:2023shf}. From these periods we compute the K\"ahler potential
\begin{equation}
    e^{-K} = 3 A^2 |z|^{2/3}  - \frac{54|z|}{\pi^4}  -\frac{|z|}{\pi^2}  \log|z| (\log|z|-12)  + \frac{27 |z|^{4/3}}{A^2\pi^8}+\mathcal{O}(|z|^{5/3})
\end{equation}
The terms we wrote down all come from the period vector given in \eqref{eq:Cseries}. We note that starting at order $|z|^{5/3}$ there are corrections, which are the first terms that depend on the axion $\arg z$.

\paragraph{Limiting mixed Hodge structure.} We finally characterize the limiting mixed Hodge structure arising at this singularity, which for all three conifold points at $z=\infty$ takes the shape
\begin{equation}
    \dim I^{p,q}  = \begin{tikzpicture}[baseline={([yshift=-.5ex]current bounding box.center)},scale=0.5,cm={cos(45),sin(45),-sin(45),cos(45),(15,0)}]
\draw (0,0) node {$0$};
\draw (1,0) node {$0$};
\draw (2,0) node {$0$};
\draw (3,0) node {$0$};
\draw (4,0) node {$1$};

\draw (0,1) node {$0$};
\draw (1,1) node {$1$};
\draw (2,1) node {$0$};
\draw (3,1) node {$0$};
\draw (4,1) node {$0$};

\draw (0,2) node {$0$};
\draw (1,2) node {$0$};
\draw (2,2) node {$1$};
\draw (3,2) node {$0$};
\draw (4,2) node {$0$};

\draw (0,3) node {$0$};
\draw (1,3) node {$0$};
\draw (2,3) node {$0$};
\draw (3,3) node {$1$};
\draw (4,3) node {$0$};

\draw (0,4) node {$1$};
\draw (1,4) node {$0$};
\draw (2,4) node {$0$};
\draw (3,4) node {$0$};
\draw (4,4) node {$0$};
\end{tikzpicture}\,  .
\end{equation}
For the example at hand these vector spaces are spanned by
\begin{equation}
    I^{4,0}: \left(0,0,0,1,\tfrac{1}{2}+\tfrac{i \sqrt{3}}{2} \right)\, , \qquad I^{3,3}: \left(1,0,\tfrac{1}{2},0,0 \right)\, ,
\end{equation}
and $I^{0,4}$ follows by complex conjugation. $I^{2,2}$ and $I^{1,1}$ are obtained through
\begin{equation}
    N_\infty = \log M_u = \left(
\begin{array}{ccccc}
 0 & 0 & 0 & 0 & 0 \\
 1 & 0 & 0 & 0 & 0 \\
 \tfrac{1}{2} & -1 & 0 & 0 & 0 \\
 0 & 0 & 0 & 0 & 0 \\
 0 & 0 & 0 & 0 & 0 \\
\end{array}
\right)\, ,
\end{equation}
by acting on $I^{3,3}$. Again $M_{ss}$ is an automorphism of this limiting mixed Hodge structure: $I^{4,0}$ has eigenvalue $e^{2\pi i/3}$, $I^{3,3},I^{2,2},I^{1,1}$ eigenvalue $-1$, and $I^{0,4}$ eigenvalue $e^{4\pi i/3}$. Also the expansion terms in the period vector \eqref{eq:Cseries} can be identified with particular subspaces.

\subsection{F-points}
Finally we consider the case where the singularity at $z=\infty$ has an F-point, i.e.~a point with finite order monodromy. These singular points are characterized by local exponents $a_1,\ldots, a_5$ that are all distinct. Among the fourfolds in table \ref{table:data} there are seven such cases, and we take \#8 -- the mirror sextic -- as working example here. This case has local exponents $(\tfrac{1}{6},\tfrac{1}{3},\tfrac{1}{2},\tfrac{2}{3},\tfrac{5}{6})$. 

\paragraph{Monodromy matrix and pairing.} We first simplify the expressions for the boundary data by applying the unimodular basis transformation
\begin{equation}
    B = \scalebox{0.9}{$\left(
\begin{array}{ccccc}
 1 & 3 & -3 & 3 & 4 \\
 0 & -1 & 2 & -1 & -2 \\
 0 & 0 & 1 & 0 & -1 \\
 1 & 3 & -7 & 2 & 8 \\
 0 & 1 & -3 & 0 & 4 \\
\end{array}
\right)$} \, .
\end{equation}
In this basis the pairing and monodromy matrix read
\begin{equation}\label{eq:Fmonodromy}
    \eta = \scalebox{0.9}{$\left(
\begin{array}{ccccc}
 0 & 0 & -1 & -1 & 2 \\
 0 & 0 & 1 & -2 & 2 \\
 -1 & 1 & -4 & 2 & 0 \\
 -1 & -2 & 2 & -4 & 0 \\
 2 & 2 & 0 & 0 & 6 \\
\end{array}
\right)$}\, , \qquad M_\infty = \scalebox{0.9}{$\left(
\begin{array}{ccccc}
 0 & 1 & 0 & 0 & 0 \\
 -1 & 1 & 0 & 0 & 0 \\
 0 & 0 & -1 & 1 & 0 \\
 0 & -1 & -1 & 0 & 0 \\
 0 & -1 & 0 & 0 & -1 \\
\end{array}
\right)$}\, .
\end{equation}
Notice that on the diagonal of $M_\infty$ we have $2\times 2$ subblocks of $\bbZ_3$ and $\bbZ_6$ monodromies of $SL(2,\bbZ)$. However, we also have a non-trivial mixing between these blocks due to non-vanishing off-diagonal components that cannot be removed by integral basis transformations.

\paragraph{Asymptotic periods.} To describe the periods close to the F-point we perform the coordinate transformation
\begin{equation}
    z \to - 6^{-3} z^{-1}\, ,
\end{equation}
Note that, in contrast to the other three types of singularities, the rescaling by $6^{-3}$ here is not fixed by canceling any logarithms, as we have a finite order monodromy. Instead, we just chose it such that the coefficients in the expansion of the period vector work out in a convenient way. The leading expansion of the period vector then reads
\begin{equation}\label{eq:Fperiods}
\begin{aligned}
    \mathbf\Pi &= A z^{1/6}\left(
\begin{array}{c}
 1 \\
 \frac{1}{2}+\frac{i \sqrt{3}}{2} \\
 -\frac{1}{2} \\
 -\frac{3}{4}-\frac{i \sqrt{3}}{4} \\
 -\frac{1}{2}-\frac{i}{2 \sqrt{3}} \\
\end{array}
\right) + B z^{1/3} \left(
\begin{array}{c}
 0 \\
 0 \\
 1 \\
 \frac{1}{2}+\frac{i \sqrt{3}}{2} \\
 0 \\
\end{array}
\right)  -\frac{\sqrt{\frac{3}{2}}z^{1/2}}{4 \pi ^2} \left(
\begin{array}{c}
 0 \\
 0 \\
 0 \\
 0 \\
 1 \\
\end{array}
\right) \\
& \ \ \ + \frac{3 z^{2/3}}{16 \pi^4 B}\left(
\begin{array}{c}
 0 \\
 0 \\
 1 \\
 \frac{1}{2}-\frac{i \sqrt{3}}{2} \\
 0 \\
\end{array}
\right) + \frac{z^{5/6}}{16\pi^4 A}\left(
\begin{array}{c}
 1 \\
 \frac{1}{2}-\frac{i \sqrt{3}}{2} \\
 -\frac{1}{2} \\
 -\frac{3}{4}+\frac{i \sqrt{3}}{4} \\
 -\frac{1}{2}+\frac{i}{2 \sqrt{3}} \\
\end{array}
\right) + \mathcal{O}(z^{7/6})\, .
\end{aligned}
\end{equation}
where we defined the non-algebraic coefficients
\begin{equation}
    A = -\frac{1458 i \sqrt{2} \Gamma \left(\frac{7}{6}\right)^6}{\pi ^5} \simeq -4.29559 i\, , \qquad B = -\frac{3 \sqrt{3} \Gamma \left(\frac{1}{3}\right)^6}{32 \pi ^5} \simeq -0.196137\, .
\end{equation}
From these periods we obtain as K\"ahler potential
\begin{equation}
    e^{-K} = \frac{1}{2} |A|^2 |z|^{1/3} - 6 B^2 |z|^{2/3} + \frac{9}{16\pi^4}|z| - \frac{3 |z|^{4/3}}{128 \pi^8 B^2 } - \frac{9 |z|^{5/3}}{512 \pi^8|A|^2} + \mathcal{O}(|z|^{4/3})
\end{equation}
We included all terms obtained from the periods \eqref{eq:Fperiods}, but note that there are other corrections appearing at lower orders: the piece at order $z^{1/6}$ in $\mathbf{\Pi}$ combines with the term at order $\bar z^{7/6}$ in $\mathbf{\bar \Pi}$ to give a term at order $|z|^{4/3}$. Moreover, this term breaks the continuous shift symmetry of the axion $\arg z$ into a discrete symmetry.

\paragraph{Boundary Hodge structure.} At the F-point the limiting mixed Hodge structure is given simply by a Hodge structure spanned by
\begin{equation}\label{eq:HpqF}
\begin{aligned}
    H^{4,0}_\infty &: \ \left(1,\tfrac{1}{2}+\tfrac{i \sqrt{3}}{2},-\tfrac{1}{2},-\tfrac{3}{4}-\tfrac{i \sqrt{3}}{4},-\tfrac{1}{2}-\tfrac{i}{2 \sqrt{3}}\right)\, , \\
     H^{3,1}_\infty &: \ (0,0,1,\tfrac{1}{2} + \tfrac{i\sqrt{3}}{2},0) \, , \qquad H^{2,2}_\infty :\  (0,0,0,0,1)\, ,
\end{aligned}
\end{equation}
and $H^{1,3}_\infty$ and $H^{0,4}_\infty$ follow by complex conjugation. These are precisely the eigenvectors of the $M_\infty$ given in \eqref{eq:Fmonodromy}, and also give the expansion terms of the period vector in \eqref{eq:Fperiods}.

\paragraph{Moduli stabilization.} While at the other three types of singularities the Calabi--Yau manifold degenerates, it does not at an F-point, so we can consider flux vacua located there. The F-term equation requires $G_4$ to have vanishing pairing with $H^{3,1}_\infty$, which yields
\begin{equation}
    \mathbf{G}_4 = ( -2g_3, 2g_3-2g_4, g_3, g_4,g_5)\, .
\end{equation}
Computing the pairing with the vector spanning $H^{4,0}_\infty$ gives as vacuum superpotential
\begin{equation}
    |W_0|^2 =  e^{K}|W|^2 = \frac{2}{3}(3 g_3^2 -3 g_3 g_4 + g_4^2)\, .
\end{equation}
Having $W_0=0$ thus requires $g_3=g_4=0$. Such a vacuum can be found for each of the seven F-points, as this four-form flux is given by the $(-1)$-eigenvector of $M_\infty$, which is always integer. More general methods for finding flux vacua with $W=0$ at special loci appear in the upcoming work \cite{toappear}. For the general case $g_3,g_4 \neq 0$ the tadpole contribution from fluxes reads
\begin{equation}
    N_{\rm flux} = \langle G_4, G_4 \rangle = 4 (3 g_3^2 -3 g_3 g_4 + g_4^2) -8 g_4 g_5 + 6 g_5^2\, ,
\end{equation}
which reduces to $N_{\rm flux}=6 g_5^2$ for flux vacua with a vanishing superpotential. It would be interesting to search for flux vacua also away from these special points, as was recently done in \cite{Plauschinn:2023hjw} for Type IIB orientifolds on the mirror octic.

\section{Conclusions}\label{sec:conclusions}
In this work we have classified Calabi--Yau fourfolds whose complex structure moduli space is the thrice-punctured sphere $\cM_{\rm cs} = \bbP^1\backslash\{0,1,\infty\}$. Our main result is that --- under certain assumptions about two of the singularity types --- we find only 14 such Calabi--Yau fourfolds. 

This classification was based on the monodromies around the punctures. Our assumption was that at the first two punctures we have large complex structure and conifold points. Their monodromy matrices are parametrized by two integers: the intersection number $\kappa$ and integrated second Chern class $c_2$ of the mirror Calabi--Yau fourfold. In turn, demanding the remaining monodromy around infinity $M_\infty$ to be quasi-unipotent permits only 14 possible values for these numbers, as summarized in table \ref{table:monodromydata}.

The Picard-Fuchs equation for the periods is a hypergeometric differential equation, with its parameters fixed by the eigenvalues of $M_\infty$. The fundamental solution at the large complex structure point can be expressed as a hypergeometric series with factorial coefficients. From this combinatorial data we read off the mirror Calabi--Yau fourfolds as complete intersections in weighted projective spaces, giving us five new cases in addition to the nine geometries of \cite{CaboBizet:2014ovf}. There are two ways to bring the solutions for the periods into an integral basis. The first way is to demand the monodromy transformations to match with our monodromy matrices, which fixes the transition matrix uniquely. A second equivalent way is to use the K-theory basis of \cite{Gerhardus:2016iot, Cota:2017aal, Marchesano:2021gyv} fixed by the topological data of the mirror Calabi--Yau fourfold. All data required to study these models has been summarized in table \ref{table:data}, and an accompanying notebook is included that computes the period vector for any of these.

The Calabi--Yau fourfolds we found are in one-to-one correspondence to the 14 Calabi--Yau threefolds of \cite{Doran:2005gu, almkvist2005tables}. The most straightforward way to match them up is by looking at the exponents of the periods at infinity, where we simply need to add another $\frac{1}{2}$ to relate them. This relation carries over to the fundamental periods and thereby also to the mirror CICYs. On the other hand, for the Calabi--Yau fourfolds themselves there is a nested fibration structure going down to lower-dimensional Calabi--Yau manifolds as described in \cite{Doran:2015xjb}. In particular, all fourfolds are elliptic fibrations over $\bbP^1\times \bbP^1\times \bbP^1$, justifying their use for F-theory compactifications.

This landscape features four different types  of phases at $z=\infty$: an LCS point, a CY3-point, a conifold point, and a LG point. We presented a working example for each of these phases, providing characteristic data such as the asymptotic periods and the K\"ahler potential. From these models we have drawn some preliminary insights, such as at what order the continuous shift symmetry of axions is broken. We also found closed expressions for certain couplings such as D7-brane flux superpotentials, whose meaning in the Type IIB picture remains to be understood. Nonetheless, further exploration of these moduli spaces, including their potential phenomenological applications, still provides an intriguing avenue for future research.

Another natural next step is to implement the strategy used in this work to other corners of the landscape. For instance, can we extend to the multi-moduli case, or forego the assumed large complex structure and conifold monodromies?  Calabi--Yau threefolds without large complex structure points are known to exist, cf.~\cite{rohde2009maximal, cynk2017picard}, so it would be exciting to get a handle on this kind of Calabi--Yau fourfold as well. The finiteness theorem of \cite{DeligneFiniteness} already tells us that there can only be finitely many possibilities given a moduli space and its singularity structure, but devising effective methods for enumerating all monodromy groups requires us to improve our understanding of the underlying duality structures.

\subsection*{Acknowledgements}
It is a pleasure to thank Chuck Doran, Naomi Gendler, Thomas Grimm, Christoph Nega, Severin L\"ust, Thorsten Schimannek, John Stout, and Max Wiesner for useful discussions. This work is supported in part by a grant from the Simons Foundation (602883, CV), the Della Pietra Foundation, by the NSF grant PHY-2013858, and NSF grant PHY-2309135 to the Kavli Institute for Theoretical Physics (KITP).

\appendix
\addtocontents{toc}{\protect\setcounter{tocdepth}{1}}

\section{Relation to K-theory basis}\label{app:Ktheory}
In this appendix we relate our integral basis to the K-theory basis used in \cite{Gerhardus:2016iot, Cota:2017aal, Marchesano:2021gyv}. The latter basis was constructed by using results from K-theory for the integral periods near the LCS point. Here we explain how the two are related by an integral basis transformation. 

\paragraph{LCS periods in K-theory basis.} Let us begin by writing down the large complex structure periods in the K-theory basis. We parametrize this regime by the covering coordinate $t= \log[z]/2\pi i$, such that the LCS point $z=0$ is mapped to $t=i\infty$. From \cite{Cota:2017aal} we then recall that the periods in the K-theory basis to leading order read
\begin{equation}
    \mathbf{\tilde{\Pi}} = \left(
\begin{array}{c}
 1 \\
 -t \\
 \frac{\kappa  t^2}{2}+\kappa  t+\frac{7 \kappa }{12}+\frac{c_2}{24} \\
 -\frac{\kappa  t^3}{6}-\frac{\kappa 
   t^2}{4}-\frac{\kappa  t}{6}-\frac{c_2 t}{24}-\frac{\kappa }{24}-\frac{c_2}{48}+\frac{i c_3 \zeta (3)}{8 \pi ^3} \\
 \frac{\kappa  t^4}{24}+\frac{c_2 t^2}{48}-\frac{i c_3 t \zeta (3)}{8 \pi ^3}-\frac{c_4}{3456}+\frac{7}{12} \\
\end{array}
\right) + \mathcal{O}(e^{2\pi i t})\, ,
\end{equation}
where we use the tilde to indicate that we use the K-theory basis.  The bilinear pairing of $SO(3,2)$ in this basis is given by
\begin{equation}
    \tilde{\eta} = \begin{pmatrix}
        0 & 0 & 0 & 0 & 1 \\
        0 & 0 & 0 & -1 & 0 \\
        0 & 0 & \kappa & \frac{1}{2}\kappa & \frac{1}{12}(c_2 +7 \kappa) \\
        0 & -1 & \frac{1}{2}\kappa & -\frac{1}{12}(c_2+\kappa) & -\frac{1}{24}(c_2 + \kappa) \\
        1 & 0 & \frac{1}{12}(c_2+7\kappa) & - \frac{1}{24}(c_2+ \kappa) & 2
    \end{pmatrix}
\end{equation}
Note that all entries are integral by the quantization condition \eqref{eq:quantization} and even $\kappa$. A similar form for the intersection pairing has been found in \cite{Marchesano:2021gyv}, although there are some convention differences. We may straightforwardly read of the monodromy matrix around the large complex structure point $t \to t+1$ as well as
\begin{equation}
    \tilde{M}_0 = \left(
\begin{array}{ccccc}
 1 & 0 & 0 & 0 & 0 \\
 -1 & 1 & 0 & 0 & 0 \\
 \frac{3 \kappa }{2} & -\kappa  & 1 & 0 & 0 \\
 0 & 0 & -1 & 1 & 0 \\
 0 & 0 & 0 & -1 & 1 \\
\end{array}
\right)\, ,
\end{equation}
Note that it satisfies $(\tilde{M}_0)^T \eta^{-1} \tilde{M}_0$, as the K-theory basis is a homology basis.

\paragraph{Our LCS periods.} Let us next write down our periods \eqref{eq:piLCS} in a homology basis, which requires a multiplication with the pairing \eqref{eq:eta}. Since we want to preserve the lower-triangular form of the monodromies, we additionally reverse the entries, which yields
\begin{equation}
    \sigma \mathbf{\Pi} = \begin{pmatrix}
        1 \\
        -t \\
        -\frac{\kappa}{2}t^2 -\frac{\kappa}{2}t + \frac{c_2}{24} \\
        \frac{\kappa}{6}t^3 + \frac{\kappa}{4}t^2 + \frac{\kappa}{8}t -\frac{c_2}{48} + \frac{c_3 \zeta(3)}{(2\pi i)^3} \\
        \frac{\kappa}{24}t^4 + \frac{c_2 }{48}t^2 + \frac{c_3 t \zeta(3)}{(2\pi i)^3}- \frac{5}{12}-\frac{c_4}{3456}
    \end{pmatrix}\, ,
\end{equation}
where we defined $\sigma =\text{diag}(1,1,\kappa,1,1)$.

\paragraph{Basis transformation.} We compute the rotation matrix back to the original basis for the periods in \eqref{eq:piLCS} to be
\begin{equation}
    B = \left(
\begin{array}{ccccc}
 1 & 0 & 0 & 0 & 0 \\
 0 & 1 & 0 & 0 & 0 \\
 \frac{1}{12}(7\kappa+c_2) & -\frac{\kappa }{2} & -1 & 0 & 0 \\
 -\frac{1}{24} (\kappa +c_2) & \frac{1}{24}(\kappa +c_2) & 0 & -1 & 0 \\
 1 & 0 & 0 & 0 & 1 \\
\end{array}
\right) \in GL(5,\bbZ)\, , \qquad \sigma \mathbf\Pi = B \mathbf{\tilde\Pi}\, .
\end{equation}
Note that this basis transformation is integral by the quantization conditions \eqref{eq:quantization} on the topological data. This basis transformation relates the monodromy matrices as
\begin{equation}
    \tilde{\eta} = B^{-1} \eta B^{-1,T} \, , \qquad \tilde{M}_0 = B \sigma M_0 \sigma^{-1}B^{-1} \, ,
\end{equation}
where the extra rescaling $\sigma$ is needed to bring the monodromy matrix to the homology basis.

\section{Overview of Calabi--Yau threefolds}\label{app:threefolds}
In this appendix we briefly review hypergeometric period systems of Calabi--Yau threefolds \cite{Doran:2005gu, almkvist2005tables}. We describe the differential equation for the periods and explain how to fix the integral basis. In table \ref{table:hypergeom} we summarized the relevant data for all 14 models. We also included a notebook detailing the period computations for any of these models.

\begin{table}[h!]
\begin{center}
	\begin{tabular}{|c|c|c|c|c|c|c|}
		\hline
	       $a_1,a_2,a_3,a_4$	& Type & $\mu$	& Mirror 		& $\kappa$	& $c_2 $	& $c_3$  \\ 
\hline & & & & & &       \\[-1.5ex]
 $\frac{1}{5},\frac{2}{5},\frac{3}{5},\frac{4}{5}$			& F & $5^5$		& $X_5(1^5)$			& $5$		& $50$		& $-200$\\[2mm]
 $\frac{1}{10},\frac{3}{10},\frac{7}{10},\frac{9}{10}$			& F & $2^85^5$	& $X_{10}(1^3 2^1 5^1)$		& $1$		& $34$		& $-288$\\[2mm]
 $\frac{1}{2},\frac{1}{2},\frac{1}{2},\frac{1}{2}$		&LCS 	& $2^8$		& $X_{2,2,2,2}(1^8)$		& $16$		& $64$		& $-128$\\[2mm]
 $\frac{1}{3},\frac{1}{3},\frac{2}{3},\frac{2}{3}$			& K& $3^6$		& $X_{3,3}(1^6)$		& $9$		& $54$		& $-144$\\[2mm]
 $\frac{1}{3},\frac{1}{2},\frac{1}{2},\frac{2}{3}$			& C& $2^43^3$	& $X_{3,2,2}(1^7)$		& $12$		& $60$		& $-144$\\[2mm]
 $\frac{1}{4},\frac{1}{2},\frac{1}{2},\frac{3}{4}$			& C &$2^{10}$	& $X_{4,2}(1^6)$		& $8$		& $56$		& $-176$\\[2mm]
 $\frac{1}{8},\frac{3}{8},\frac{5}{8},\frac{7}{8}$			& F & $2^{16}$	& $X_{8}(1^4 4^1)$		& $2$		& $44$		& $-296$\\[2mm]
 $\frac{1}{6},\frac{1}{3},\frac{2}{3},\frac{5}{6}$			& F & $2^43^6$	& $X_{6}(1^4 2^1)$			& $3$		& $42$		& $-204$\\[2mm]
 $\frac{1}{12},\frac{5}{12},\frac{7}{12},\frac{11}{12}$		& F & $2^{12}3^6$	& $X_{12,2}(1^4 4^1 6^1)$	& $1$		& $46$		& $-484$\\ [ 2 mm]
 $\frac{1}{4},\frac{1}{4},\frac{3}{4},\frac{3}{4}$			& K& $2^{12}$	& $X_{4,4}(1^4 2^2)$		& $4$		& $40$		& $-144$\\[2mm]
 $\frac{1}{4},\frac{1}{3},\frac{2}{3},\frac{3}{4}$			& F  & $2^63^3$	& $X_{4,3}(1^5 2^1)$		& $6$		& $48$		& $-156$\\[2mm]
 $\frac{1}{6},\frac{1}{4},\frac{3}{4},\frac{5}{6}$			& F & $2^{10}3^3$	& $X_{6,4}(1^3 2^2 3^1)$	& $2$	& $32$ & $-156$\\[2mm]
 $\frac{1}{6},\frac{1}{6},\frac{5}{6},\frac{5}{6}$			& K& $2^83^6$	& $X_{6,6}(1^2 2^2 3^2)$		& $1$		& $22$		& $-120$\\[2mm]
$\frac{1}{6},\frac{1}{2},\frac{1}{2},\frac{5}{6}$			& C&  $2^83^3$	& $X_{6,2}(1^5 3^1)$		& $4$		& $52$		& $-256$\\[2mm]
		\hline
	 \end{tabular}	
\end{center}
\caption{\label{table:hypergeom}Data of the 14 Calabi--Yau threefolds with $\cM_{\rm cs}=\bbP^1\backslash \{0,1,\infty\}$. }
\end{table}

\paragraph{Picard-Fuchs equation.} As was also the case for elliptic curves and Calabi--Yau fourfolds, the Picard-Fuchs equation for the periods is given by the hypergeometric operator
\begin{equation}
    L = \theta^4 - \mu z \prod_{i=1}^4 (\theta+a_i)\, .
\end{equation}
The position of the conifold singularity is fixed as
\begin{equation}
    \mu^{-1} = e^{4\gamma_E+\sum_{i=1}^4\psi(a_i)}\, ,
\end{equation}
where $\gamma_E$ is the Euler-Mascheroni constant and $\psi(x)$ the digamma function. This choice removes all $\log \mu$ terms from the periods in the large complex structure regime.

\paragraph{Frobenius solution.} The Frobenius ansatz for the large complex structure point reads
\begin{align}
    \varpi_0 &= f_0(z)\, ,\nonumber \\
    2\pi i\, \varpi_1 &= f_0(z) \log[z]+f_1(z)\, , \nonumber \\
    (2\pi i)^2\,\varpi_2 &= f_0(z) \log[z]^2 + 2 f_1(z) \log[z]+ f_2(z) \, , \nonumber \\
    (2\pi i)^3\,\varpi_3 &= f_0(z) \log[z]^3 + 3 f_1(z) \log[z]^2+ 3 f_2(z) \log[z]+f_3(z)\, , 
\end{align}
where we defined the holomorphic power series
\begin{equation}
    f_i(z) = \delta_{i,0}+\sum_{k=1}^\infty c_{i,k}z^k\, .
\end{equation}
By plugging these power series into the Picard-Fuchs equation, one can solve for the coefficients $c_{i,k}$ order-by-order. We can perform a similar expansion around the conifold point $z=\mu^{-1}$. Shifting $z\to z+\mu^{-1}$ such that it is located at $z=0$, we write the series expansions
\begin{equation}
\begin{aligned}
       \varpi_0 &=   f_0(z)\, , \quad  & \varpi_{1} &=   z f_1(z)\, , \quad &2\pi i \, \varpi_{2,1} &=   z f_1(z) \log[z]+z f_2(z)\, , \quad &\varpi_3 &=   z^2f_3 (z)\, ,\\
\end{aligned}
\end{equation}
where all power series are given by
\begin{equation}
f_i(z) = 1+\sum_{k=1}^\infty c_{i,k}z^k\, \text{ for $i=0,1,3$}\, , \qquad f_2(z) = \sum_{k=1}^\infty c_{2,k}z^k\, .
\end{equation}
Here we note that by linearly combining solutions we may set $c_{0,1}=c_{0,2}=0$ and $c_{2,2}=0$, such that at next-to-leading order we have $f_0(z)=1+c_{0,3}z^3$ and $f_2(z)=c_{2,3} x^2$ respectively. On the other hand, we note that $c_{1,2}\neq 0$; even though it is arbitrary for the Picard-Fuchs equation of $\varpi_1$, it is needed to take a particular value for $\varpi_2$ in order to solve its differential equation.

\paragraph{Transition matrix.} The transition matrix from the Frobenius basis to the integral basis at the large complex structure point can be given in terms of the mirror topological data as
\begin{equation}\label{eq:CY3T}
    \cT_{\rm LCS} = \begin{pmatrix}
        1 & 0 & 0 & 0 \\
        0 & 1 & 0 & 0 \\
        \frac{c_2}{24} & \sigma & -\kappa/2 & 0 \\
        -\frac{c_3 \zeta(3)}{(2\pi i)^3} & \frac{c_2}{24} & 0 & \kappa/6
    \end{pmatrix}\, .
\end{equation}
where $\kappa$ is the intersection number, $c_2,c_3$ are integrated Chern classes, and $\sigma=\kappa/2 \mod 1$.

\paragraph{Topological data.} While the topological data is readily listed in table \ref{table:hypergeom}, it may also be obtained from just the exponents $a_1,\ldots,a_4$. The intersection number follows from
\begin{equation}
    \kappa = 2^4 \prod_i \sin(a_i)\, ,
\end{equation}
while the Chern classes may be computed as
\begin{equation}
    c_2 = \kappa \Big(-2+\frac{3}{\pi^2} \sum_{i=1}^4 \psi_1(a_i) \Big) \, , \qquad c_3 = \frac{\kappa}{6}\Big(8+\frac{1}{\zeta(3)}\sum \psi_2(a_i) \big)\, .
\end{equation}

\section{Limiting mixed Hodge structures}
\label{app:lmhs}
Here we briefly review some concepts about limiting mixed Hodge structures. We refer the reader to \cite{vandeHeisteeg:2022gsp} for a more detailed review on asymptotic Hodge theory and its application to string compactifications.

\paragraph{Monodromy weight filtration.} In order to define the limiting mixed Hodge structure, we first introduce the \textit{monodromy weight filtration} associated to $N = \log T_u$. This is a set of rational vector spaces $W_\ell(N)$ (with $\ell=0,\ldots,8$) that forms an increasing filtration $W_\ell \subset W_{\ell+1}$, defined as
\begin{equation}
    W_\ell(N) = \sum_{j \geq \rm{max}(-1,\ell-4)} \ker N^{j+1} \cap \text{img}\, N^{j-\ell+4} \, ,
\end{equation}
in terms of the kernels and images of $N$.

\paragraph{Limiting filtration.} In addition to the monodromy weight filtration, we also need the limiting Hodge filtration: a set of complex vector spaces $F^p_0$ (with $p=0,\ldots, 4$) forming a decreasing filtration $F^p_0 \subseteq F^{p-1}_0$. It is obtained from the Hodge filtration 
\begin{equation}
    F^p = \text{span}\{ \mathbf\Pi \, , \ \ldots \, , \  \partial^{4-p} \mathbf\Pi\}\, ,
\end{equation}
by taking the approximation close to the boundary as
\begin{equation}
    F^p(z) \simeq e^{\frac{\log z}{2\pi i} N } F^p_0\, ,
\end{equation}
where the vector spaces $F^p_0$ are independent of $z$. In practice this means we need to keep as many terms in the period vector $\mathbf\Pi$ as we need to span a $(4-p)$-dimensional vector space $F^p_0$.

\paragraph{Limiting mixed Hodge structures.} With this monodromy weight filtration $W_\ell(N)$ and limiting filtration $F^p_0$ in place, we define the limiting mixed Hodge structure by the splitting
\begin{equation}
    I^{p,q} = F^p_0 \cap \overline{F^q_0} \cap W_{p+q}\, .
\end{equation}
In the general expression $\overline{F^p_0}\cap W_{p+q}$ is replaced by a sum over vector spaces. In this work the additional terms are not needed, as we can perform a convenient change of coordinates that removes them.

\paragraph{Semi-simple monodromies.} The semisimple monodromy factors $M_{ss}$ of finite order act as automorphisms on the limiting mixed Hodge structure. In general this means that they act on elements spanning these vector spaces as
\begin{equation}
    \omega_{p,q} \in I^{p,q}: \qquad M_{ss} \omega_{p,q} \in I^{p,q}\, .
\end{equation}
In the present case we can be even more precise, since the eigenvalues of the semisimple factor $M_{ss}$ are fixed by the exponents $a_i$ in the Picard-Fuchs equation \eqref{eq:pfgen} as $e^{2\pi i a_i}$. Then the action of the semisimple factor is given by
\begin{equation}\label{eq:MssIpq}
    \omega_{p,q} \in I^{p,q}: \qquad M_{ss} \omega_{p,q} = e^{2\pi i a_p} \omega_{p,q}\, ,
\end{equation}
as one may verify explicitly for the working examples presented in section \ref{sec:infmonodromies}.

\bibliographystyle{JHEP}
\bibliography{refs}

\providecommand{\href}[2]{#2}\begingroup\raggedright\begin{thebibliography}{10}

\bibitem{Vafa:2005ui}
C.~Vafa, {\it {The String landscape and the swampland}},
  \href{http://arxiv.org/abs/hep-th/0509212}{{\tt hep-th/0509212}}.

\bibitem{Palti:2019pca}
E.~Palti, {\it {The Swampland: Introduction and Review}},  {\em Fortsch. Phys.}
  {\bf 67} (2019), no.~6 1900037, [\href{http://arxiv.org/abs/1903.06239}{{\tt
  arXiv:1903.06239}}].

\bibitem{vanBeest:2021lhn}
M.~van Beest, J.~Calder\'on-Infante, D.~Mirfendereski, and I.~Valenzuela, {\it
  {Lectures on the Swampland Program in String Compactifications}},  {\em Phys.
  Rept.} {\bf 989} (2022) 1--50, [\href{http://arxiv.org/abs/2102.01111}{{\tt
  arXiv:2102.01111}}].

\bibitem{Agmon:2022thq}
N.~B. Agmon, A.~Bedroya, M.~J. Kang, and C.~Vafa, {\it {Lectures on the string
  landscape and the Swampland}},  \href{http://arxiv.org/abs/2212.06187}{{\tt
  arXiv:2212.06187}}.

\bibitem{Ooguri:2006in}
H.~Ooguri and C.~Vafa, {\it {On the Geometry of the String Landscape and the
  Swampland}},  {\em Nucl. Phys. B} {\bf 766} (2007) 21--33,
  [\href{http://arxiv.org/abs/hep-th/0605264}{{\tt hep-th/0605264}}].

\bibitem{Grimm:2018ohb}
T.~W. Grimm, E.~Palti, and I.~Valenzuela, {\it {Infinite Distances in Field
  Space and Massless Towers of States}},  {\em JHEP} {\bf 08} (2018) 143,
  [\href{http://arxiv.org/abs/1802.08264}{{\tt arXiv:1802.08264}}].

\bibitem{Blumenhagen:2018nts}
R.~Blumenhagen, D.~Klaewer, L.~Schlechter, and F.~Wolf, {\it {The Refined
  Swampland Distance Conjecture in Calabi-Yau Moduli Spaces}},  {\em JHEP} {\bf
  06} (2018) 052, [\href{http://arxiv.org/abs/1803.04989}{{\tt
  arXiv:1803.04989}}].

\bibitem{Lee:2018urn}
S.-J. Lee, W.~Lerche, and T.~Weigand, {\it {Tensionless Strings and the Weak
  Gravity Conjecture}},  {\em JHEP} {\bf 10} (2018) 164,
  [\href{http://arxiv.org/abs/1808.05958}{{\tt arXiv:1808.05958}}].

\bibitem{Grimm:2018cpv}
T.~W. Grimm, C.~Li, and E.~Palti, {\it {Infinite Distance Networks in Field
  Space and Charge Orbits}},  {\em JHEP} {\bf 03} (2019) 016,
  [\href{http://arxiv.org/abs/1811.02571}{{\tt arXiv:1811.02571}}].

\bibitem{Corvilain:2018lgw}
P.~Corvilain, T.~W. Grimm, and I.~Valenzuela, {\it {The Swampland Distance
  Conjecture for K\"ahler moduli}},  {\em JHEP} {\bf 08} (2019) 075,
  [\href{http://arxiv.org/abs/1812.07548}{{\tt arXiv:1812.07548}}].

\bibitem{Joshi:2019nzi}
A.~Joshi and A.~Klemm, {\it {Swampland Distance Conjecture for One-Parameter
  Calabi-Yau Threefolds}},  {\em JHEP} {\bf 08} (2019) 086,
  [\href{http://arxiv.org/abs/1903.00596}{{\tt arXiv:1903.00596}}].

\bibitem{Font:2019cxq}
A.~Font, A.~Herr\'aez, and L.~E. Ib\'a\~nez, {\it {The Swampland Distance
  Conjecture and Towers of Tensionless Branes}},  {\em JHEP} {\bf 08} (2019)
  044, [\href{http://arxiv.org/abs/1904.05379}{{\tt arXiv:1904.05379}}].

\bibitem{Marchesano:2019ifh}
F.~Marchesano and M.~Wiesner, {\it {Instantons and infinite distances}},  {\em
  JHEP} {\bf 08} (2019) 088, [\href{http://arxiv.org/abs/1904.04848}{{\tt
  arXiv:1904.04848}}].

\bibitem{Grimm:2019wtx}
T.~W. Grimm and D.~Van De~Heisteeg, {\it {Infinite Distances and the Axion Weak
  Gravity Conjecture}},  {\em JHEP} {\bf 03} (2020) 020,
  [\href{http://arxiv.org/abs/1905.00901}{{\tt arXiv:1905.00901}}].

\bibitem{Erkinger:2019umg}
D.~Erkinger and J.~Knapp, {\it {Refined swampland distance conjecture and
  exotic hybrid Calabi-Yaus}},  {\em JHEP} {\bf 07} (2019) 029,
  [\href{http://arxiv.org/abs/1905.05225}{{\tt arXiv:1905.05225}}].

\bibitem{Lee:2019oct}
S.-J. Lee, W.~Lerche, and T.~Weigand, {\it {Emergent strings from infinite
  distance limits}},  {\em JHEP} {\bf 02} (2022) 190,
  [\href{http://arxiv.org/abs/1910.01135}{{\tt arXiv:1910.01135}}].

\bibitem{Baume:2019sry}
F.~Baume, F.~Marchesano, and M.~Wiesner, {\it {Instanton Corrections and
  Emergent Strings}},  {\em JHEP} {\bf 04} (2020) 174,
  [\href{http://arxiv.org/abs/1912.02218}{{\tt arXiv:1912.02218}}].

\bibitem{Klawer:2021ltm}
D.~Kl\"awer, {\it {Modular curves and the refined distance conjecture}},  {\em
  JHEP} {\bf 12} (2021) 088, [\href{http://arxiv.org/abs/2108.00021}{{\tt
  arXiv:2108.00021}}].

\bibitem{Alvarez-Garcia:2021mzv}
R.~\'Alvarez-Garc\'\i{}a and L.~Schlechter, {\it {Analytic periods via twisted
  symmetric squares}},  {\em JHEP} {\bf 07} (2022) 024,
  [\href{http://arxiv.org/abs/2110.02962}{{\tt arXiv:2110.02962}}].

\bibitem{Alvarez-Garcia:2021pxo}
R.~\'Alvarez-Garc\'\i{}a, D.~Kl\"awer, and T.~Weigand, {\it {Membrane limits in
  quantum gravity}},  {\em Phys. Rev. D} {\bf 105} (2022), no.~6 066024,
  [\href{http://arxiv.org/abs/2112.09136}{{\tt arXiv:2112.09136}}].

\bibitem{Candelas:1990rm}
P.~Candelas, X.~C. De~La~Ossa, P.~S. Green, and L.~Parkes, {\it {A Pair of
  Calabi-Yau manifolds as an exactly soluble superconformal theory}},  {\em
  Nucl. Phys. B} {\bf 359} (1991) 21--74.

\bibitem{gross1993finiteness}
M.~Gross, {\it A finiteness theorem for elliptic calabi-yau threefolds},
  \href{http://arxiv.org/abs/alg-geom/9305002}{{\tt alg-geom/9305002}}.

\bibitem{wilson2023boundedness}
P.~M.~H. Wilson, {\it Boundedness questions for calabi-yau threefolds},
  \href{http://arxiv.org/abs/1706.01268}{{\tt arXiv:1706.01268}}.

\bibitem{Kreuzer:2000xy}
M.~Kreuzer and H.~Skarke, {\it {Complete classification of reflexive polyhedra
  in four-dimensions}},  {\em Adv. Theor. Math. Phys.} {\bf 4} (2002)
  1209--1230, [\href{http://arxiv.org/abs/hep-th/0002240}{{\tt
  hep-th/0002240}}].

\bibitem{Gendler:2023ujl}
N.~Gendler, N.~MacFadden, L.~McAllister, J.~Moritz, R.~Nally, A.~Schachner, and
  M.~Stillman, {\it {Counting Calabi-Yau Threefolds}},
  \href{http://arxiv.org/abs/2310.06820}{{\tt arXiv:2310.06820}}.

\bibitem{Chandra:2023afu}
A.~Chandra, A.~Constantin, C.~S. Fraser-Taliente, T.~R. Harvey, and A.~Lukas,
  {\it {Enumerating Calabi-Yau Manifolds: Placing bounds on the number of
  diffeomorphism classes in the Kreuzer-Skarke list}},
  \href{http://arxiv.org/abs/2310.05909}{{\tt arXiv:2310.05909}}.

\bibitem{Doran:2005gu}
C.~F. Doran and J.~W. Morgan, {\it {Mirror symmetry and integral variations of
  Hodge structure underlying one parameter families of Calabi-Yau threefolds}},
   in {\em {Workshop on Calabi-Yau Varieties and Mirror Symmetry}},
  pp.~517--537, 5, 2005.
\newblock \href{http://arxiv.org/abs/math/0505272}{{\tt math/0505272}}.

\bibitem{CaboBizet:2014ovf}
N.~Cabo~Bizet, A.~Klemm, and D.~Vieira~Lopes, {\it {Landscaping with fluxes and
  the E8 Yukawa Point in F-theory}},
  \href{http://arxiv.org/abs/1404.7645}{{\tt arXiv:1404.7645}}.

\bibitem{Lee:2021qkx}
S.-J. Lee and T.~Weigand, {\it {Elliptic K3 surfaces at infinite complex
  structure and their refined Kulikov models}},  {\em JHEP} {\bf 09} (2022)
  143, [\href{http://arxiv.org/abs/2112.07682}{{\tt arXiv:2112.07682}}].

\bibitem{Lee:2021usk}
S.-J. Lee, W.~Lerche, and T.~Weigand, {\it {Physics of infinite complex
  structure limits in eight dimensions}},  {\em JHEP} {\bf 06} (2022) 042,
  [\href{http://arxiv.org/abs/2112.08385}{{\tt arXiv:2112.08385}}].

\bibitem{Alvarez-Garcia:2023gdd}
R.~\'Alvarez-Garc\'\i{}a, S.-J. Lee, and T.~Weigand, {\it {Non-minimal Elliptic
  Threefolds at Infinite Distance I: Log Calabi-Yau Resolutions}},
  \href{http://arxiv.org/abs/2310.07761}{{\tt arXiv:2310.07761}}.

\bibitem{Alvarez-Garcia:2023qqj}
R.~\'Alvarez-Garc\'\i{}a, S.-J. Lee, and T.~Weigand, {\it {Non-minimal Elliptic
  Threefolds at Infinite Distance II: Asymptotic Physics}},
  \href{http://arxiv.org/abs/2312.11611}{{\tt arXiv:2312.11611}}.

\bibitem{Griffiths1970}
P.~A. Griffiths, {\it Periods of integrals on algebraic manifolds, iii (some
  global differential-geometric properties of the period mapping)},  {\em
  Publications Math{\'e}matiques de l'Institut des Hautes {\'E}tudes
  Scientifiques} {\bf 38} (1970) 125--180.

\bibitem{DeligneFiniteness}
P.~Deligne, {\it Un theoreme de finitude pour la monodromie},  {\em 1–19,
  Progr. Math., 67} (1987).

\bibitem{Schmid}
W.~Schmid, {\it {Variation of Hodge structure: the singularities of the period
  mapping}},  {\em Invent. Math. , 22:211--319, 1973} (1973).

\bibitem{Landman}
A.~Landman, {\it On the picard-lefschetz transformation for algebraic manifolds
  acquiring general singularities},  {\em Transactions of the American
  Mathematical Society} {\bf 181} (1973) 89--126.

\bibitem{almkvist2005tables}
G.~Almkvist, C.~van Enckevort, D.~van Straten, and W.~Zudilin, {\it Tables of
  calabi--yau equations},  {\em arXiv preprint math/0507430} (2005).

\bibitem{vanstraten2017calabiyau}
D.~van Straten, {\it Calabi--yau operators},  2017.

\bibitem{Bastian:2021hpc}
B.~Bastian, T.~W. Grimm, and D.~van~de Heisteeg, {\it {Engineering Small Flux
  Superpotentials and Mass Hierarchies}},
  \href{http://arxiv.org/abs/2108.11962}{{\tt arXiv:2108.11962}}.

\bibitem{Bastian:2023shf}
B.~Bastian, D.~van~de Heisteeg, and L.~Schlechter, {\it {Beyond Large Complex
  Structure: Quantized Periods and Boundary Data for One-Modulus
  Singularities}},  \href{http://arxiv.org/abs/2306.01059}{{\tt
  arXiv:2306.01059}}.

\bibitem{Haack:2001jz}
M.~Haack and J.~Louis, {\it {M theory compactified on Calabi-Yau fourfolds with
  background flux}},  {\em Phys. Lett. B} {\bf 507} (2001) 296--304,
  [\href{http://arxiv.org/abs/hep-th/0103068}{{\tt hep-th/0103068}}].

\bibitem{Denef:2008wq}
F.~Denef, {\it {Les Houches Lectures on Constructing String Vacua}},  {\em Les
  Houches} {\bf 87} (2008) 483--610,
  [\href{http://arxiv.org/abs/0803.1194}{{\tt arXiv:0803.1194}}].

\bibitem{Grimm:2010ks}
T.~W. Grimm, {\it {The N=1 effective action of F-theory compactifications}},
  {\em Nucl. Phys. B} {\bf 845} (2011) 48--92,
  [\href{http://arxiv.org/abs/1008.4133}{{\tt arXiv:1008.4133}}].

\bibitem{Weigand:2018rez}
T.~Weigand, {\it {F-theory}},  {\em PoS} {\bf TASI2017} (2018) 016,
  [\href{http://arxiv.org/abs/1806.01854}{{\tt arXiv:1806.01854}}].

\bibitem{Gukov:1999ya}
S.~Gukov, C.~Vafa, and E.~Witten, {\it {CFT's from Calabi-Yau four folds}},
  {\em Nucl. Phys. B} {\bf 584} (2000) 69--108,
  [\href{http://arxiv.org/abs/hep-th/9906070}{{\tt hep-th/9906070}}]. [Erratum:
  Nucl.Phys.B 608, 477--478 (2001)].

\bibitem{GKZ}
I.~M. Gel'fand, A.~V. Zelevinskii, and M.~Kapranov, {\it Hypergeometric
  functions and toral manifolds},  {\em Functional Analysis and Its
  Applications} {\bf 23} (1989) 94--106.

\bibitem{Hosono:1993qy}
S.~Hosono, A.~Klemm, S.~Theisen, and S.-T. Yau, {\it {Mirror symmetry, mirror
  map and applications to Calabi-Yau hypersurfaces}},  {\em Commun. Math.
  Phys.} {\bf 167} (1995) 301--350,
  [\href{http://arxiv.org/abs/hep-th/9308122}{{\tt hep-th/9308122}}].

\bibitem{Hosono:1994ax}
S.~Hosono, A.~Klemm, S.~Theisen, and S.-T. Yau, {\it {Mirror symmetry, mirror
  map and applications to complete intersection Calabi-Yau spaces}},  {\em
  Nucl. Phys. B} {\bf 433} (1995) 501--554,
  [\href{http://arxiv.org/abs/hep-th/9406055}{{\tt hep-th/9406055}}].

\bibitem{Norland}
N.~E. N{\o}rlund, {\it {Hypergeometric functions}},  {\em Acta Mathematica}
  {\bf 94} (1955), no.~none 289 -- 349.

\bibitem{Gerhardus:2016iot}
A.~Gerhardus and H.~Jockers, {\it {Quantum periods of Calabi\textendash{}Yau
  fourfolds}},  {\em Nucl. Phys. B} {\bf 913} (2016) 425--474,
  [\href{http://arxiv.org/abs/1604.05325}{{\tt arXiv:1604.05325}}].

\bibitem{MirandaPersson}
R.~Miranda and U.~Persson, {\it On extremal rational elliptic surfaces},  {\em
  Mathematische Zeitschrift} {\bf 193} (1986), no.~4 537--558.

\bibitem{Doran:2015xjb}
C.~F. Doran and A.~Malmendier, {\it {Calabi\textendash{}Yau manifolds realizing
  symplectically rigid monodromy tuples}},  {\em Adv. Theor. Math. Phys.} {\bf
  23} (2019), no.~5 1271--1359, [\href{http://arxiv.org/abs/1503.07500}{{\tt
  arXiv:1503.07500}}].

\bibitem{Schimannek:2021pau}
T.~Schimannek, {\it {Modular curves, the Tate-Shafarevich group and
  Gopakumar-Vafa invariants with discrete charges}},  {\em JHEP} {\bf 02}
  (2022) 007, [\href{http://arxiv.org/abs/2108.09311}{{\tt arXiv:2108.09311}}].

\bibitem{Zagier}
D.~Zagier, {\it Integral solutions of apéry-like recurrence equations},  {\em
  Harnad, John; Winternitz, Pavel: Groups and symmetries, American Mathematical
  Society, 349-366 (2009)} {\bf 47} (07, 2009).

\bibitem{Grimm:2009ef}
T.~W. Grimm, T.-W. Ha, A.~Klemm, and D.~Klevers, {\it {Computing Brane and Flux
  Superpotentials in F-theory Compactifications}},  {\em JHEP} {\bf 04} (2010)
  015, [\href{http://arxiv.org/abs/0909.2025}{{\tt arXiv:0909.2025}}].

\bibitem{Seidel:2000ia}
P.~Seidel and R.~P. Thomas, {\it {Braid group actions on derived categories of
  coherent sheaves}},  \href{http://arxiv.org/abs/math/0001043}{{\tt
  math/0001043}}.

\bibitem{Cota:2017aal}
C.~F. Cota, A.~Klemm, and T.~Schimannek, {\it {Modular Amplitudes and
  Flux-Superpotentials on elliptic Calabi-Yau fourfolds}},  {\em JHEP} {\bf 01}
  (2018) 086, [\href{http://arxiv.org/abs/1709.02820}{{\tt arXiv:1709.02820}}].

\bibitem{Marchesano:2021gyv}
F.~Marchesano, D.~Prieto, and M.~Wiesner, {\it {F-theory flux vacua at large
  complex structure}},  {\em JHEP} {\bf 08} (2021) 077,
  [\href{http://arxiv.org/abs/2105.09326}{{\tt arXiv:2105.09326}}].

\bibitem{Klemm:2004km}
A.~Klemm, M.~Kreuzer, E.~Riegler, and E.~Scheidegger, {\it {Topological string
  amplitudes, complete intersection Calabi-Yau spaces and threshold
  corrections}},  {\em JHEP} {\bf 05} (2005) 023,
  [\href{http://arxiv.org/abs/hep-th/0410018}{{\tt hep-th/0410018}}].

\bibitem{clingher201614th}
A.~Clingher, C.~F. Doran, J.~Lewis, A.~Y. Novoseltsev, and A.~Thompson, {\it
  The 14th case vhs via k3 fibrations},  {\em Recent advances in Hodge theory:
  Period domains, algebraic cycles and arithmetic, London Math. Soc. Lecture
  Note Ser} {\bf 427} (2016) 165--227.

\bibitem{Grimm:2019ixq}
T.~W. Grimm, C.~Li, and I.~Valenzuela, {\it {Asymptotic Flux Compactifications
  and the Swampland}},  {\em JHEP} {\bf 06} (2020) 009,
  [\href{http://arxiv.org/abs/1910.09549}{{\tt arXiv:1910.09549}}]. [Erratum:
  JHEP 01, 007 (2021)].

\bibitem{Katz:2022lyl}
S.~Katz, A.~Klemm, T.~Schimannek, and E.~Sharpe, {\it {Topological Strings on
  Non-Commutative Resolutions}},  \href{http://arxiv.org/abs/2212.08655}{{\tt
  arXiv:2212.08655}}.

\bibitem{Doran:2016uea}
C.~F. Doran, A.~Harder, and A.~Thompson, {\it {Mirror symmetry, Tyurin
  degenerations and fibrations on Calabi-Yau manifolds}},  {\em Proc. Symp.
  Pure Math.} {\bf 96} (2017) 93--132,
  [\href{http://arxiv.org/abs/1601.08110}{{\tt arXiv:1601.08110}}].

\bibitem{tyurin}
A.~N. Tyurin, {\it Fano versus calabi - yau},  2003.

\bibitem{Katzarkov}
L.~Katzarkov and V.~Przyjalkowski, {\it Generalized homological mirror symmetry
  and cubics},  {\em Proceedings of the Steklov Institute of Mathematics} {\bf
  264} (2009), no.~1 87--95.

\bibitem{Sen:1996vd}
A.~Sen, {\it {F theory and orientifolds}},  {\em Nucl. Phys.} {\bf B475} (1996)
  562--578, [\href{http://arxiv.org/abs/hep-th/9605150}{{\tt hep-th/9605150}}].

\bibitem{Clingher:2012rg}
A.~Clingher, R.~Donagi, and M.~Wijnholt, {\it {The Sen Limit}},  {\em Adv.
  Theor. Math. Phys.} {\bf 18} (2014), no.~3 613--658,
  [\href{http://arxiv.org/abs/1212.4505}{{\tt arXiv:1212.4505}}].

\bibitem{Alim:2009bx}
M.~Alim, M.~Hecht, H.~Jockers, P.~Mayr, A.~Mertens, and M.~Soroush, {\it {Hints
  for Off-Shell Mirror Symmetry in type II/F-theory Compactifications}},  {\em
  Nucl. Phys. B} {\bf 841} (2010) 303--338,
  [\href{http://arxiv.org/abs/0909.1842}{{\tt arXiv:0909.1842}}].

\bibitem{Jockers:2009ti}
H.~Jockers, P.~Mayr, and J.~Walcher, {\it {On N=1 4d Effective Couplings for
  F-theory and Heterotic Vacua}},  {\em Adv. Theor. Math. Phys.} {\bf 14}
  (2010), no.~5 1433--1514, [\href{http://arxiv.org/abs/0912.3265}{{\tt
  arXiv:0912.3265}}].

\bibitem{Bonisch:2022mgw}
K.~B\"onisch, A.~Klemm, E.~Scheidegger, and D.~Zagier, {\it {D-brane masses at
  special fibres of hypergeometric families of Calabi-Yau threefolds, modular
  forms, and periods}},  \href{http://arxiv.org/abs/2203.09426}{{\tt
  arXiv:2203.09426}}.

\bibitem{toappear}
T.~W. Grimm and D.~van~de Heisteeg, {\it {to appear}}, .

\bibitem{Plauschinn:2023hjw}
E.~Plauschinn and L.~Schlechter, {\it {Flux vacua of the mirror octic}},  {\em
  JHEP} {\bf 01} (2024) 157, [\href{http://arxiv.org/abs/2310.06040}{{\tt
  arXiv:2310.06040}}].

\bibitem{rohde2009maximal}
J.~C. Rohde, {\it Maximal automorphisms of calabi-yau manifolds versus
  maximally unipotent monodromy},  2009.

\bibitem{cynk2017picard}
S.~Cynk and D.~van Straten, {\it Picard-fuchs operators for octic arrangements
  i (the case of orphans)},  {\em arXiv preprint arXiv:1709.09752} (2017).

\bibitem{vandeHeisteeg:2022gsp}
D.~T.~E. van~de Heisteeg, {\em {Asymptotic String Compactifications: Periods,
  flux potentials, and the swampland}}.
\newblock PhD thesis, Utrecht U., 2022.
\newblock \href{http://arxiv.org/abs/2207.00303}{{\tt arXiv:2207.00303}}.

\end{thebibliography}\endgroup

\end{document}